\documentclass[twocolumn]{aastex61}
\usepackage[utf8]{inputenc}
\usepackage{graphicx}
\usepackage{hyperref}
\usepackage{amsmath}
\usepackage{amssymb}
\usepackage{tabularx}
\usepackage{multirow}
\usepackage{bm}
\usepackage{hhline}

\newcommand{\GG}[1]{}

\usepackage{float}
\usepackage{placeins}

\begin{document}

\title{\textit{Galaxy Zoo: Clump Scout}: Surveying the Local Universe for Giant Star-forming Clumps}

\author{Dominic Adams} \affiliation{School of Physics and Astronomy, University of Minnesota, 116 Church Street SE, Minneapolis, MN 55455, US}

\author{Vihang Mehta} \affiliation{School of Physics and Astronomy, University of Minnesota, 116 Church Street SE, Minneapolis, MN 55455, US}

\author{Hugh Dickinson} \affiliation{School of Physical Sciences, The Open University, Walton Hall, MK7 6AA Milton Keynes, UK}

\author{Claudia Scarlata} \affiliation{School of Physics and Astronomy, University of Minnesota, 116 Church Street SE, Minneapolis, MN 55455, US}

\author{Lucy Fortson} \affiliation{School of Physics and Astronomy, University of Minnesota, 116 Church Street SE, Minneapolis, MN 55455, US}

\author{Sandor Kruk} \affiliation{European Space Agency, ESTEC, Keplerlaan 1, NL-2201 AZ, Noordwijk, The Netherlands} \affiliation{Max-Planck-Institut für extraterrestrische Physik (MPE),Giessenbachstrasse 1, D-85748 Garching bei München, Germany}

\author{Brooke Simmons} \affiliation{Physics Department, Lancaster University, Lancaster, LA1 4YB, UK}

\author{Chris Lintott} \affiliation{Oxford Astrophysics, Denys Wilkinson Building, Keble Road, Oxford, OX1 3RH, UK}

\begin{abstract}
    Massive, star-forming clumps are a common feature of high-redshift star-forming galaxies. How they formed, and why they are so rare at low redshift, remains unclear. In this paper we identify the largest yet sample of clumpy galaxies (7,052) at low redshift using data from the citizen science project \textit{Galaxy Zoo: Clump Scout}, in which volunteers classified over 58,000 Sloan Digital Sky Survey (SDSS) galaxies spanning redshift $0.02 < z < 0.15$. We apply a robust completeness correction by comparing with simulated clumps identified by the same method. Requiring that the ratio of clump-to-galaxy flux in the SDSS $u$ band be greater than 8\% (similar to clump definitions used by other works), we estimate the fraction of local galaxies hosting at least one clump ($f_{clumpy}$) to be $2.68_{-0.30}^{+0.33}\%$. We also compute the same fraction with a less stringent cut of 3\% ($11.33_{-1.16}^{+0.89}\%$), as the higher number count and lower statistical noise of this fraction permits sharper comparison with future low-redshift clumpy galaxy studies. Our results reveal a sharp decline in $f_{clumpy}$ over $0 < z < 0.5$. The minor merger rate remains roughly constant over the same span, so we suggest that minor mergers are unlikely to be the primary driver of clump formation. Instead, the rate of galaxy turbulence is a better tracer for $f_{clumpy}$ over $0 < z < 1.5$ for galaxies of all masses, which supports the idea that clump formation is primarily driven by violent disk instability for all galaxy populations during this period.
\end{abstract}

\keywords{Starburst galaxies (1570), Galaxies (573), Galaxy evolution (594), Galaxy formation (595), Galaxy structure (622), Cosmological evolution (336)}

\section{Introduction}

The morphologies of low-redshift galaxies can generally be classified by their location on the Hubble sequence \citep{Hubble1926}. However, in the last two decades, observations of star-forming galaxies at the peak of cosmic star formation ($z\sim2$) reveal that they typically exhibit irregular, clumpy morphologies different from these classifications \citep{CowieHuSongaila1995, ElmegreenElmegreenHirst2004a, ElmegreenElmegreenSheets2004b}. Clumpy galaxies receive their name from the ``giant star-forming clumps'' that occupy them. \citet{Elmegreen2007} estimated these clumps to be of characteristic mass between $10^{7-9} M_\odot$ in galaxies in the Hubble Ultra Deep Field, much more massive than typical star-forming regions locally (though more recent papers have suggested lower characteristic masses, e.g. \cite{Fisher+2017, DessuagesZavadskyAdamo2018}). It has now been established that clumpy morphology in star-forming galaxies peaks near $z\sim2$, with $\gtrsim 50\%$ exhibiting clumpy behavior, before declining significantly with cosmic time \citep{Murata+2014, Guo+2015, Shibuya+2016, Guo+2018}. 

It is still not clear why clumps are so dominant at $z \sim 2$, nor why they are so much rarer in low-redshift galaxies. Two major modes of clump formation have been discussed at length in the literature. First, clumps may form ``in-situ'' due to ``violent disk instability'' (VDI) within their host galaxy, i.e. in disks where the Toomre stability parameter is $\sim 1$ \citep{Bournaud+2014, Mandelker+2014, Fisher+2017}. Such disks are called ``marginally stable''. This state can occur in galaxies that are continuously fed by smooth cold accretion of intergalactic gas \citep{Genzel+2008, DekelSariCeverino2009}, though observational confirmation of this process has been limited \citep{Scarlata+2009, Bouche+2013}. Second, clumps may form due to interactions, i.e. major or minor mergers. The merging galaxies can themselves become clumps, known as ``ex-situ'' clumps, or they can generate local instabilities within the merging galaxies which collapse to form in-situ clumps. It is expected that ex-situ clumps are higher in mass and volume, older, and lower in star-formation activity than their in-situ counterparts; simulations and observations both suggest that a minority of high-redshift clumps are formed this way \citep{Mandelker+2014, Zanella+2019}. It is possible that different formation mechanisms are at work for different populations of clumpy galaxies: \cite{Guo+2015} suggests that the clump formation mode depends strongly on galaxy mass, with high-mass galaxies ($M \gtrsim 10^{10.5} M_\odot$) dominated by the VDI-driven ``in-situ'' mode while low-mass galaxies ($M \lesssim 10^{10} M_\odot$) are dominated by minor-merger-driven ``ex-situ'' formation.

Clumps have been difficult to study largely because they are difficult to resolve with existing instruments. While clumps were originally thought to be kiloparsec-scale objects, high-resolution observations of clumpy galaxies in the local universe \citep[e.g.][]{Overzier+2009, Fisher+2014, Fisher+2017, Messa+2019} and in strongly lensed fields \citep{Wuyts+2014, DessuagesZavadsky+2017, DessuagesZavadskyAdamo2018, Cava+2018} have revealed that many clumps have a much smaller characteristic size, ranging from tens to hundreds of parsecs. It has therefore been theorized that high-redshift ``kiloparsec-scale clumps'' may be rare, and that the size and mass of observed clumps are an order of magnitude lower than were originally estimated. Several papers point to ``blending'' of multiple small clumps at low resolution to explain high-redshift observations \citep{DessuagesZavadsky+2017, Fisher+2017, DessuagesZavadskyAdamo2018}. The characteristic size, mass and properties of giant star-forming clumps is still a topic of debate.

To date, studies of local clumpy galaxies have focused on high-resolution imaging of small galaxy samples ($n < 50$). However, no studies have yet assembled a large catalog of local clumpy galaxies. Local clumpy galaxies, while rare, are observable in much greater detail and can act as analogs to their more distant counterparts. To this end, we launched the citizen science project \textit{Galaxy Zoo: Clump Scout} (herein called \textit{Clump Scout}). This project, which was active between 2019 and 2021, recruited volunteers to visually identify clumps in galaxies in the local universe. Each subject was examined by many volunteers, whose annotations were aggregated into consensus locations. The catalog of low-redshift clumps and clumpy galaxies from this project is the largest of its kind, and can help to constrain models of clump formation and galaxy evolution by comparing it with high-redshift populations.

In this paper we will present the clump catalog assembled by \textit{Clump Scout} and use it to estimate the fraction of clumpy galaxies ($f_{clumpy}$) in the local universe. \cite{Guo+2015}, \cite{Shibuya+2016} and others have used the evolution of $f_{clumpy}$ with redshift to evaluate the likelihood of different clump formation mechanisms (i.e. by internal VDI or by galaxy interactions). We will extend their analysis using our own $f_{clumpy}$ result.

This paper is structured as follows. Section 2 describes the galaxy sample, the \textit{Clump Scout} citizen science project, and our methods for aggregating the annotations provided by \textit{Clump Scout} volunteers. Section 3 describes our methods for recovering clump properties and correcting for incompleteness. Section 4 describes the criteria applied to define clumps, and also details the clump catalog accompanying this paper. Section 5 estimates the local fraction of clumpy galaxies and compares it to other works. In Section 6 we discuss this result, its physical significance, and potential issues. Section 7 presents a summary and conclusions.

\section{Sample selection and preparation}

\subsection{The \textit{Galaxy Zoo: Clump Scout} project}

\textit{Galaxy Zoo: Clump Scout} was a citizen science project which was active from September 19, 2019 to February 11, 2021 on the Zooniverse platform\footnote{\url{https://z.umn.edu/ClumpScout}}. It presented volunteers with image cutouts of galaxies from the Sloan Digital Sky Survey \citep[SDSS;][]{York+2000_SDSS_legacy}; for each galaxy image (referred to as a subject) the volunteer was asked to identify its central bulge, followed by all of its off-center clumps. The central bulge location was requested in order to discourage volunteers from marking the galaxy center as a clump, as well as to remove any identified ``clumps'' coinciding with galaxy center. To improve classification quality, new volunteers were presented with a classification tutorial; many examples of correct classifications were provided by a field guide and task-level help menus as well. In addition, the majority of volunteers\footnote{This feature could only be made available for volunteers who were logged into the platform. Roughly 83\% of classifications were provided by logged-in volunteers.} were shown a small sample of expert-classified ``training'' images when they began classifying ($<10$ randomly interspersed throughout their first 20 classifications). After classifying a training image, the volunteer was given feedback on how the classification compared to that of the experts.

\begin{figure*}
    \centering
    \includegraphics[width=\textwidth]{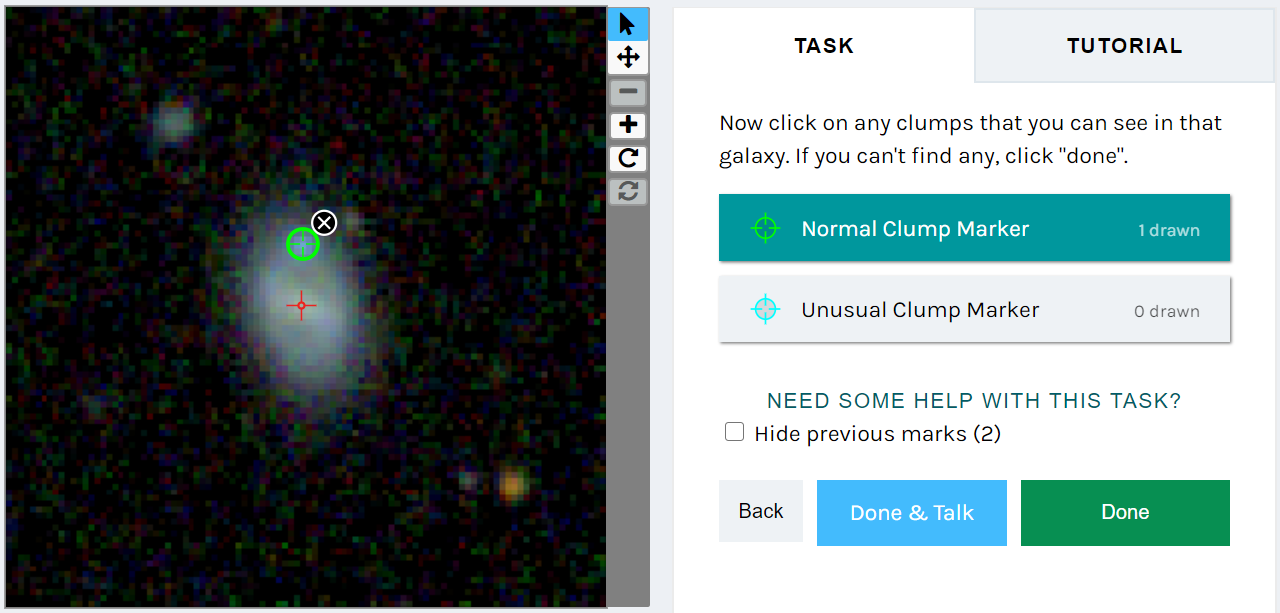}
    \caption{The user interface for \textit{Galaxy Zoo: Clump Scout}, showing a partially completed classification. The volunteer has marked the central bulge (red crosshairs) in the previous step and is now identifying clumps.}
    \label{fig:clump_scout_interface}
\end{figure*}

For the clump-marking task, volunteers were provided with a ``normal clump marker'' and an ``unusual clump marker'' (see Figure \ref{fig:clump_scout_interface}). The ``unusual'' marker was intended to identify foreground star contaminants, i.e. Milky Way stars that overlap with the angular area of the target galaxy. Like clumps, foreground stars appear as point sources in SDSS images, and their colors in the $i$, $r$ and $g$ bands can be very similar to clumps; this makes them very difficult to distinguish from clumps by any simple rule. We therefore instructed volunteers to mark clumps as ``unusual'' if they were particularly bright, differently-colored, or especially offset from their host galaxy. Examples of ``unusual clumps'' (which were generally foreground stars) were provided in the tutorial and field guide.

\subsection{Galaxy selection criteria}
\label{sec:galaxy_selection}

\begin{table}[]
    \centering
    \begin{tabular}{|l|l|}
        \hline
        \multicolumn{1}{|c|}{\textbf{Selection}} & \multicolumn{1}{c|}{\textbf{Galaxy count}} \\ \hhline{|=|=|}
        \multicolumn{2}{|c|}{\textbf{Parent sample}}  \\ \hline
        GZ2 with spec. redshift & 243,500 \\ \hline
        With MPA-JHU mass estimates & 239,950 \\ \hline
        With $z>0.02$ & 225,085 \\ \hhline{|=|=|}
        \multicolumn{2}{|c|}{\textbf{Regular sample}}  \\ \hline        
        With $f_{merger}<0.5$ & 211,236 \\ \hline
        With $f_{featured} > 0.5$
        & \textbf{53,613} \color{black} \\ \hhline{|=|=|}
        \multicolumn{2}{|c|}{\textbf{Extra sample}} \\ \hline
        With $f_{featured} \geq 0.5$ or $f_{merger} \geq 0.5$ & 171,472 \\ \hline
        With $z < 0.075$ & 75,832 \\ \hline
        Randomly selected sample & \textbf{4,937} \\ \hline
        
    \end{tabular}
    \caption{The sequence of cuts performed on the galaxy sample. $f_{merger}$ refers to the debiased GZ2 vote fraction for ``merger'', while $f_{featured}$ refers to the weighted GZ2 vote fraction for ``features or disk''. The final sample consisted of all 53,613 galaxies with $f_{featured} > 0.5$ (the ``regular'' sample), in addition to a sample of 4,937 of the 171,472 galaxies with $f_{merger} \geq 0.5$ and $f_{featured} \leq 0.5$ (the ``extra'' sample).}
    \label{tab:sample_selection}
\end{table}

In this study, our goal was to identify as thorough a sample of clumpy galaxies in the local Universe as possible, where we take ``local'' to refer to a region nearby enough that little to no cosmological evolution takes place (i.e. $z \lesssim 0.15$). To select for potentially clumpy galaxies, we relied on data from the \textit{Galaxy Zoo 2} (GZ2) citizen science project, which provides morphological classifications for over 300,000 local galaxies from the SDSS legacy survey. GZ2 itself consists of roughly 25\% of galaxies identified in the SDSS main galaxy sample, and comprises ``the nearest, brightest, and largest systems for which fine morphological features can be resolved and classified'' \cite{Willett+2013}. From 243,500 GZ2 galaxies with spectroscopic redshifts, we selected for galaxies with detectable features by requiring a weighted vote fraction\footnote{The fraction of volunteers who voted for a particular \textit{Galaxy Zoo 2} classification, weighted according the estimated consistency of each volunteer.} of at least 50\% for the presence of ``features or disk'' ($f_{featured}$). Major mergers were removed by requiring that the debiased vote fraction\footnote{Similar to the weighted vote fraction, but additionally corrected to remove systematic biases in galaxy properties due to their redshift.} for the presence of ``merger'' ($f_{merger}$) was below 50\%. \citep[For an overview of GZ2 vote fractions, see][]{Willett+2013}. Galaxies with $z<0.02$ were removed to ensure that the vast majority of clumps could not be resolved and would appear as point sources; this was required so that realistic-looking simulated clumps could be added to these galaxies for comparison. Finally, we limited our sample to galaxies whose masses had been estimated by the SDSS DR7 MPA-JHU value-added catalog\footnote{\url{http://www.sdss3.org/dr8/spectro/galspec.php}} \citep{Kauffmann+2003_MPA-JHU, Brinchmann+2004_MPA-JHU}. Our of GZ2 as our parent catalog means that we do not make any cuts on galaxy size, as GZ2 already selects its galaxies to be resolvable (\texttt{petroR90\_r} $>$ 3 arcsec, where \texttt{petroR90\_r} is the radius containing 90\% of galaxy flux in the Petrosian aperture). The resulting sample contained 53,613 galaxies.

In addition to this main sample, we included a sample of 4,937 additional galaxies from GZ2 with $0.02 < z < 0.075$ and for which the GZ2 weighted vote fraction for ``features or disk'' was less than 50\%. The redshift limit of $z < 0.075$ was applied to include more low-mass galaxies in our sample, since SDSS is dominated by high-mass galaxies at higher redshifts. In total, 157,623 galaxies in the GZ2 spectroscopic sample met the ``features or disk'' vote fraction cut ($f_{featured} > 0.5$), and 74,057 of these additionally met the $z < 0.075$ redshift cut. We selected only a subsample of these to study as they were not classified as having features or a disk, and therefore unlikely candidates for hosting clumps. The 4,937 galaxies chosen will be referred to as the ``extra'' sample, and we included the sample to permit us to extrapolate its clumpy statistics over the full population. With the ``extra'' sample included, a total of 58,550 galaxies were studied. Unless otherwise specified, photometry describing galaxies in our sample were obtained from the SDSS DR15 PhotoPrimary table \citep{SDSS_DR15}. Table \ref{tab:sample_selection} summarizes our galaxy selection criteria and number counts.

Given this sample, we created image cutouts for all galaxies from the SDSS DR15 Legacy survey which were presented to volunteers. We mapped the SDSS $i$, $r$, and $g$ band imaging to the red, green and blue values of the image cutouts using \textit{asinh} color scaling \citep{Lupton+2004}, and scaled galaxies to be approximately the same size in each cutout. Cutout creation is described more fully in Appendix \ref{app:cutout_creation}.

\subsection{Generating images with simulated clumps} \label{sec:sims}

\begin{figure}[h]
    \centering
    \includegraphics[width=\columnwidth]{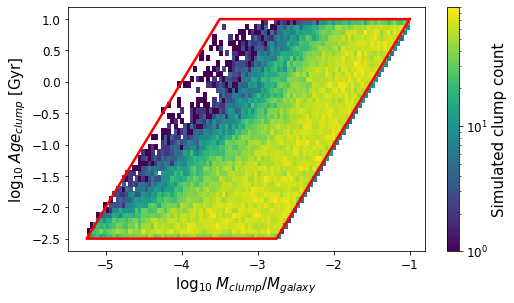}
    \caption{A 2D histogram of simulated clumps counts in log$_{10}$ clump relative mass and age space. Clumps' properties were drawn uniformly from within the area plotted in red, then discarded if they did not meet the magnitude limit described in Section \ref{sec:sims}.}
    \label{fig:sim_mass_age}
\end{figure}

\begin{figure*}[t]
    \centering
    \includegraphics[width=1\textwidth]{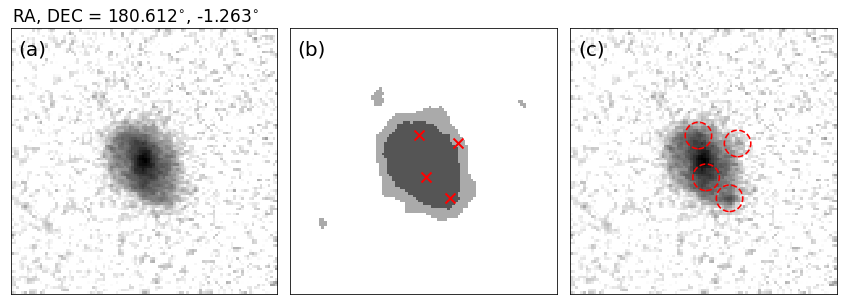}
    \caption{An illustration of the process of placing simulated clumps in an image. \textbf{(a)} The initial galaxy image, $g$ band only. \textbf{(b)} The galaxy segmentation map. The dark-gray inner region is the ``narrow'' region where clumps are placed with 75\% probability, while the dark- and light-gray regions together make up the ``wide'' region where clumps are placed with 25\% probability. Selected locations for simulated clumps are plotted as x's. \textbf{(c)} The final galaxy image with simulated clumps added (some are too faint to be detected). Dashed red lines are drawn around the simulations to highlight them. The final galaxy image shown to volunteers was rotated and/or reflected with respect to this one to distinguish its appearance from the original.}
    \label{fig:sim_placement}
\end{figure*}

To estimate the completeness of our sample, we created additional cutouts of galaxies from the target sample with artificial clumps added. These will be called ``simulated'' subjects, whereas cutouts with no added clumps will be called ``real''. The simulated sample consists of 84,565 simulated clumps in 26,736 galaxies, approximately half the galaxy count in the real sample. During \textit{Clump Scout}, each volunteer was shown a random sample of subjects drawn uniformly from the pool of all subjects (both real and simulated). A subject was retired (no longer shown to volunteers) after receiving 20 classifications.

Galaxies were selected for the simulated sample such that their mass distribution matched the mass distribution of clumpy galaxies in the \cite{Guo+2018} sample (1,132 galaxies spanning $0.5 < z < 3$, hereafter called Guo+2018). This lowered the characteristic mass of galaxies in the simulated sample: The Guo+2018 sample has a median mass of $10^{9.7} M_{\odot}$, compared to $10^{10.6} M_{\odot}$ for galaxies in the real sample. This was appropriate, as massive galaxies tend to be redder and smoother; visual inspection confirmed that a large fraction of star-forming clumps inserted into high-mass galaxies were easily distinguished as simulations. We allowed galaxies to appear in the simulated sample multiple times, applying a different image transformation (i.e. combination of rotations and reflections) to each cutout made of the galaxy to ensure the final subject was unique.

To determine clump luminosity and color, we simulated clump spectra using spectral templates generated by the software Flexible Stellar Population Synthesis \citep{ConroyGunnWhite2009, ConroyGunn2010}. We treated each clump as a single stellar population (with delta-function SFH), assigning it an age and a $V$ band total extinction $A_V$. Clump ages spanned $10^{-2.5}$ Gyr (i.e. $\sim$3 Myr) to $10^{1}$ Gyr. Dust extinction was given by a \cite{Calzetti+2000} attenuation curve, and the clump's $A_V$ value varied narrowly about the SDSS-estimated dust content for the host galaxy ($\sigma = 0.04$ in $\ln(A_V)$). Metallicity was fixed to its Solar value. The resulting spectrum was redshifted to match the host galaxy and integrated to determine the clump's broadband flux in each SDSS band.

Clump mass was selected uniformly from a distribution which depended upon the clump age, shown in Figure \ref{fig:sim_mass_age}. Qualitatively, we excluded low-mass, high-age clumps to avoid extremely faint (``known invisible'') objects. Similarly, we excluded the brightest clumps (high-mass, low-age) as these were ``known visible'' objects. The resulting sample probes the region where we are least certain of volunteers' recovery capability. We discarded and regenerated a clump if its apparent magnitude was $> 22.8$ in each of the $g$, $r$, $i$ and $z$ bands as well as $> 24.8$ in the $u$ band\footnote{For comparison, the SDSS point-source 95\% completeness magnitudes in the $u$, $g$, $r$, $i$, and $z$ bands are 22.0, 22.2, 22.2, 21.3, and 20.5 respectively.}, since these clumps were considered ``known invisible''. The resulting distribution in the age and mass of simulated clumps is shown in Figure \ref{fig:sim_mass_age}.

Each galaxy in the simulated sample was assigned a number of clumps selected from a Poisson distribution with mean 3, where 0 results were rejected and resampled. (Since 0's were rejected, the actual mean number of clumps per galaxy was 3.16.) This distribution was chosen to maximize the clump density per galaxy without appearing unrealistic to volunteers, and is not meant to reflect the true distribution of clumps per galaxy. To assign locations to clumps, image segmentation was performed on the $r$ band imaging field containing the host galaxy using Source Extractor \citep{BertinArnouts1996}. For each field, a centered cutout was made, smoothed with a boxcar filter, then segmented by Source Extractor to determine which pixels belonged to the host galaxy. A pixel was chosen uniformly at random from the host galaxy's segment to be the clump's location within the image. In order to probe the low surface brightness regions of a galaxy (without letting our sample be dominated by clumps in these regions), we randomly assigned 25\% of clumps to a ``wide'' segment which encompassed a larger area than the standard galaxy segment.\footnote{For both the narrow and wide segments, the Source Extractor settings \texttt{DEBLEND\_NTHRESH}=16 and \texttt{DEBLEND\_MINCONT}=0.01 were used; the only parameter changed was the detection threshold \texttt{DETECT\_THRESH}, which was set to 3 for the wide segment and 5 for the narrow. In addition, to generate the wide segment the galaxy image was convolved with a boxcar filter of size 7 pixels, while the narrow image was convolved with a boxcar filter of size 3 pixels.} Figure \ref{fig:sim_placement} contains an example of the simulation placement process. Each clump was then simulated as a point source with a $\gamma = 2.5$ Moffat profile; of the profiles we tried, it was found by visual inspection by the authors that a Moffat profile looked the most ``real'' to observers, and was most difficult to distinguish from real clumps. For most clumps, the full width at half maximum (FWHM) of this profile was equivalent to the FWHM of the $r$ band point spread function (PSF) in the image. We also allowed a small number of clumps to be ``extended'' by assigning each clump an effective physical radius, selected from a uniform distribution over the range [10 pc, 500 pc] (based on size limits observed by \cite{Messa+2019}). If this physical radius exceeded that of the PSF, this radius was used as the effective radius of the clump's profile. Approximately 19\% of simulated clumps (16,342 of 84,565) exceeded the seeing PSF size, by a median of 27\%. The distribution of simulated clump sizes allows us to probe the effect of clump size on recovery statistics, and is not intended to match the true distribution.

\subsection{Aggregation method} \label{sec:agg_method}

After subjects had been examined by \textit{Clump Scout} volunteers, the locations of clumps needed to be determined from the collected annotations on each subject. To this end, we developed an aggregation algorithm with which all volunteer annotations on each subject were transformed into consensus clump locations. Broadly, our aggregation process consisted of two steps: First, clump candidates were identified via a clustering algorithm; second, clumps marked ``unusual'' by a sufficient fraction of volunteers ($> 0.35$) were discarded, since these are likely to be foreground star contaminants.

The clustering algorithm we employed is adapted from \cite{Branson+2017} and is specialized for citizen science applications. A complete description of the algorithm can be found in Dickinson et al. (in prep). Briefly, we assign each volunteer a ``false positive probability'' (i.e. the chance that an annotation by this volunteer is not associated with any clump candidates), a ``false negative probability (i.e. the chance that the volunteer failed to mark a given clump candidate), and a ``scatter'' value which estimates the distance between an annotation and its intended target. These values inform a clustering algorithm which identifies clump candidates; in turn, the identified clump candidates update the volunteer statistics, and so on. To determine when each image has been fully classified, a ``risk'' value is computed for each one. Once the risk falls below a threshold value, clustering is no longer performed and its clump candidates are finalized. Each clump candidate is assigned a ``false positive probability'' based on the number and properties of volunteers who marked it. To improve the purity of our sample, we discard clumps with false positive probability $>0.6$. (By comparison with a sample of classifications by the authors, we found that the majority of volunteer-identified clumps with false positive probabilities larger than 0.6 did not correspond to any expert-identified clumps, while the majority of those below the threshold did.) We additionally remove any clumps that coincide with the volunteer-identified central bulge of a galaxy. To locate a galaxy's central bulge, we take the median x and y coordinates of central bulge annotations from all volunteers; we prune outlying annotations by removing any volunteer's central bulge annotation that falls more than 20 pixels from this location (in the 400x400 cutout image), and recalculate the central bulge location. We then remove any clumps within 1 PSF-FWHM of the central bulge.

In total (excluding unusual clumps), we identify 10,739 clumps over 7,052 galaxies in our sample. An additional 3,861 unusual clumps were identified; these are included in the final catalog, though it is likely that a majority are contaminating foreground stars.

\section{Clump properties and completeness} \label{sec:flux_recovery}

\begin{figure*}[ht!]
    \centering
    \includegraphics[width=0.75\textwidth]{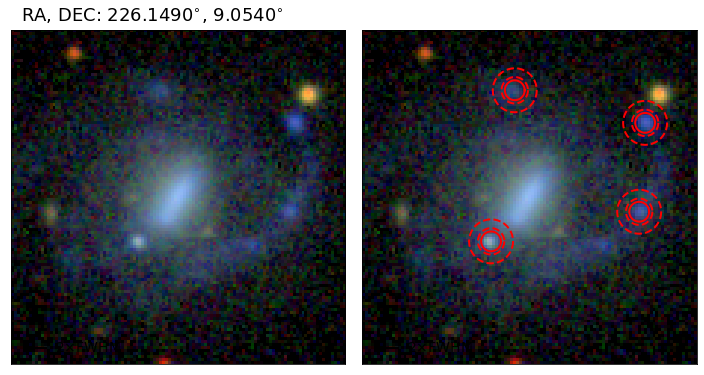}
    \caption{An illustration of the parameters used to estimate the flux from each clump. Left is a galaxy image shown to volunteers from the ``real'' sample (no simulated clumps), zoomed-in to show detail. The right image marks the locations of 4 recovered clumps. Solid outlines are drawn around the central apertures and dashed lines are drawn around the background regions.}
    \label{fig:flux_estimation}
\end{figure*}

\begin{figure*}[ht]
    \centering
    \includegraphics[width=0.8\textwidth]{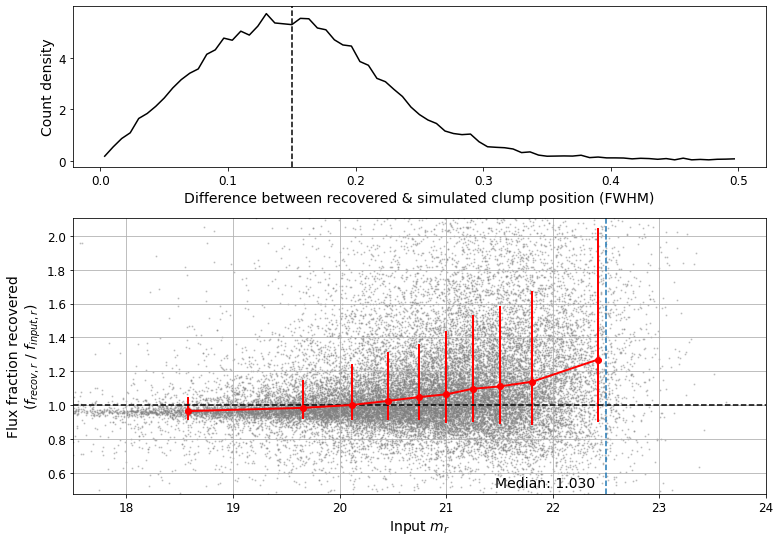}
    \caption{Here we illustrate the effectiveness of the aggregation and flux estimation process. For all simulated clumps that volunteers recovered, we compare the ``input'' properties of clumps (their original simulated values) with their ``recovered'' properties, following the aggregation process laid out in Section \ref{sec:agg_method} and the flux estimation method in Section \ref{sec:flux_recovery}. \textbf{Top:} Histogram of the offset between the recovered and true locations of simulated clumps recovered by volunteers. The median offset is 0.15 PSF-FWHM, as marked by the dashed line. \textbf{Bottom:} The flux recovery fraction (recovered flux / input flux) vs. input magnitude for simulated clumps. Red points mark the binned medians in magnitude difference, while vertical bars span the 16th to 84th percentile flux recovery fractions. The median flux recovery fraction was 1.03 for clumps with simulated magnitude $<22.5$. There are a few sources of systematic error in the flux recovery process which could not be removed completely (described in Section \ref{sec:flux_recovery}).}
    \label{fig:recovered_flux_sim}
\end{figure*}

\begin{figure*}[ht]
    \centering
    \includegraphics[width=0.85\textwidth]{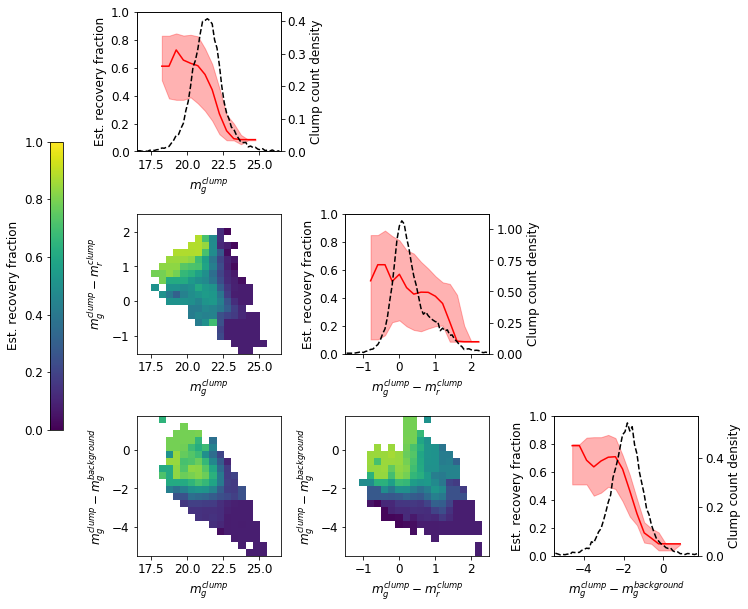}
    \caption{The estimated recovery fraction for clumps as a function of the three clump properties $m_g^{clump}$, $m_g^{clump} - m_r^{clump}$, and $m^{clump}_g - m^{background}_g$. Plots at the top of each column show the median estimated recovery fraction for clumps as a function of each property (red line), with 68\% scatter (red shaded region). Completeness estimates are not shown for bins with $<$50 clumps; the dashed black lines trace the number density of clumps as a function of each property. Plots below the diagonal show the 2D dependence of estimated recovery fraction vs. pairs of clump properties, where bins with $<$5 clumps are not shown.}
    \label{fig:completeness_estimates}
\end{figure*}

In this section, we will discuss our method for estimating the flux and background of each clump identified by \textit{Clump Scout}.

We estimated the flux of and galactic background for each clump in each of the SDSS $ugriz$ bands, using a method similar to \cite{Guo+2015}. First, flux was measured in an aperture of diameter 2.25 PSF-FWHM centered on the clump location. Next, the background-per-pixel value (where ``background'' refers to diffuse galaxy light) was estimated by taking the median pixel value in an annulus spanning diameters 3-5PSF-FWHM and used to estimate the background flux in the central aperture. Figure \ref{fig:flux_estimation} provides a visual example of the aperture and annulus sizes used. This background is subtracted to obtain a clump flux estimate within the aperture. A random sample of model PSFs from 1,000 SDSS fields revealed that $84 \pm 2$\% of flux from a point source falls within an aperture of diameter 2.25 PSF-FWHM. We therefore multiplied the background-subtracted aperture flux by 1.191 to obtain the total flux of the clump.

There are a few known sources of systematic error in our flux estimation process for simulated clumps. A small percentage of flux ($\lesssim 5\%$) is lost due to a combination of pixelation effects, contamination of the background region by the clump, and offsets between the recovered and true locations of clumps. In addition to these, the diffuse background flux is slightly underestimated by the background annulus (median: $\sim 85\%$ of the true value). This results in overestimation of clump fluxes, particularly for dimmer clumps. While these systematics cannot fully be removed with the existing method, we have minimized them so that they are smaller than the scatter. It should be noted that these systematic effects may not exactly match those for the real sample of clumps compared with our simulations; as such, we simply minimize the systematics and do not apply a correction to counteract them. The bottom panel of Figure \ref{fig:recovered_flux_sim} visually quantifies the effectiveness of this flux recovery method by displaying the recovered vs. input magnitudes of the simulated clump sample, and demonstrates that the systematic effects are less than the scatter.

To estimate the error on clump fluxes, we first estimated the per-pixel uncertainties in each SDSS field using the gain and dark variance values provided for each CCD, as well as the image calibration and sky image maps provided with each field. We then fit an uncertainty model to all the pixels in each field of the form $\sigma_f^2 = m f + b$, where $\sigma_f^2$ is the variance on a pixel's flux, $f$ is the pixel's flux, and $m$ and $b$ are model parameters. We sum the variances within a clump's aperture to estimate the aperture flux uncertainty, and we take the median pixel variance in the background annulus and multiply this by the aperture area to estimate the background uncertainty within the aperture. Finally, to obtain the variance on the background-subtracted aperture flux, we sum the aperture and annulus variance estimates; we multiply this value by the same aperture correction (1.191) to estimate the uncertainty on the final, background-subtracted clump flux.

We also obtained a completeness estimate for each clump. For a given clump, its completeness estimate is the estimated fraction of clumps similar to it that \textit{Clump Scout} recovered. To determine this we relied on the recovery fractions of simulated clumps, where a simulated clump was considered ``recovered'' if \textit{Clump Scout} volunteers located a clump within 0.75 PSF-FWHM of its location. We then examined the sample of simulated clumps with respect to three properties: A clump's brightness ($g$-band magnitude), its color ($g$-minus-$r$ magnitude), and its contrast against the diffuse background (clump-minus-background $g$-band magnitude). The simulated clumps were binned with respect to these properties, and the overall recovery fraction was calculated for each bin. We then calculated the same three properties for each real clump and compared with the recovery fractions of the simulated sample to obtain each clump's completeness estimate. The three properties selected - brightness, color and contrast - well capture the relevant properties of each clump; we found that including additional properties, including clump galactocentric radius, galaxy redshift, galaxy size, or image resolution, had only a very small effect on our completeness estimates by comparison. The details of this process are described in Appendix \ref{app:completeness_calculation}. Figure \ref{fig:completeness_estimates} shows the estimated recovery fraction statistics for real clumps in our sample.

We use these completeness estimates in this work to correct the fraction of clumpy galaxies ($f_{clumpy}$) for incompleteness (Section \ref{sec:fclumpy_completeness_corr}). Future studies relying on this clump catalog should generally incorporate these completeness estimates to accurately model the local population of clumps, and should not rely solely on the observed number counts of clumps.

\subsection{Catalog release and use} \label{sec:catalog_use}

Along with the electronic release of this paper, we release the catalog of all of the clumps identified by the Clump Scout aggregator and their estimated properties. The columns are fully described by Table \ref{tab:catalog} at the end of this paper. Clumps with a high fraction ($\geq 35\%$) of ``unusual'' annotations are included, but are marked by the flag \texttt{unusual\_flag = 1}.

To use this catalog for scientific purposes, we make a few ``best practices'' suggestions on how to filter this catalog:

\begin{itemize}
    \item \textbf{Selecting a clean sample:} We strongly recommend that clumps with \texttt{unusual\_flag = 1} should be rejected, as these are probable non-clump contaminants (i.e. foreground stars, background galaxies, or other point-like sources).
    \item \textbf{Mass completeness:} This galaxy catalog is not mass complete, and the lower-limit mass on galaxies evolves significantly with redshift. We estimate that the catalog is mass-complete down to $10^9$ M$_\odot$ for $z < 0.035$. If a mass-complete catalog is needed, care must be taken to limit the sample's redshift.
\end{itemize}

\section{The local clumpy fraction of galaxies} \label{sec:fclumpy}

A particularly important observable in the study of clumpy galaxies is the fraction of star-forming galaxies with at least one clump, known as the ``clumpy fraction'' or $f_{clumpy}$. The clumpy fraction is most simply defined
\begin{equation}
    f_{clumpy} = \frac{N(\text{SFGs with $\geq$1 clump})}{N(\text{SFGs})}
\end{equation}
where SFGs refer to star-forming galaxies (sSFR $>$ 0.1 Gyr$^{-1}$). This is an easily-compared observable between different galaxy populations which can significantly constrain models of galaxy evolution. In this section, we establish a clump definition that makes estimating $f_{clumpy}$ straightforward and precise, then present our $f_{clumpy}$ estimate.

\subsection{Selecting a clump definition for $f_{clumpy}$} \label{sec:clump_def}

A major difficulty in clumpy galaxy literature is that the definition of a ``clump'' is highly inconsistent. Past works have defined clumps as those objects identified by visual investigation \citep[e.g.][]{Elmegreen+2007, Puech2010, Overzier+2009} or by applying detection algorithms that are robust to changes in resolution and depth, including the \texttt{clumpfind} algorithm from \cite{ClumpfindPaper1994} and others \citep[e.g.][]{Livermore+2012, Guo+2012, Tadaki+2014, Zanella+2019}. However, comparisons between these different methods are not straightforward.

\cite{Guo+2015} proposed an empirically-motivated definition that the ratio of clump to galaxy UV luminosity ($f_{LUV}$) must exceed 8\%. The 8\% cutoff was chosen to select for star-forming regions at high redshift while excluding common star-forming regions locally; specifically, it includes many star-forming regions in HST-imaged galaxies spanning $0.5<z<3$, but excludes $> 99\%$ of star-forming regions identified in the galaxy M101 (blurred to match the resolution of the high-redshift sample). Local clumps exceeding the $f_{LUV} > 8\%$ threshold are therefore expected to be rare, exceptional objects. Several other recent works use this or a similar criterion \citep[e.g.][]{Shibuya+2016, Mandelker+2017, Fisher+2017}. However, it is not universally applicable: \citet{DessuagesZavadskyAdamo2018} rejected it to allow for study of the mass function of high-redshift clumps down to much lower masses, while \citet{HuertasCompany+2020} used a clump mass cut of $M_{clump} > 10^7 M_\odot$ instead to facilitate comparison with simulations.

In this paper, we calculate $f_{clumpy}$ by specifying a relative flux cutoff in the SDSS $u$ band, ie: The ratio of clump to galaxy flux in the u band ($f_{Lu}$) is greater than some specified fraction. In particular we use the relative flux cuts $f_{Lu} > 8\%$ and $f_{Lu} > 3\%$, and call the clumpy fractions under these criteria $f_{clumpy,8\%}$ and $f_{clumpy,3\%}$ respectively. The 8\% fraction was selected to be comparable to existing works with a similar criterion, while the 3\% fraction allows for larger number statistics and is easier and more accurate to estimate for the low-redshift universe where clumps are less common.

\textbf{The relationship between $f_{Lu}$ and $f_{LUV}$}: It is worth noting that past studies have defined clumps by a relative flux threshold in the UV at $\sim$2,500\AA \citep[e.g.][]{Guo+2015, Shibuya+2016}. For SDSS, UV data is not available, as the lowest wavelength band ($u$) probes $\sim$3,500\AA. However, there is a strong relationship between a clump's flux fraction in the SDSS $u$ band ($f_{Lu}$) and in the near UV ($f_{LUV}$). To demonstrate this, we examined a highly complete sample of clumps from the \cite{Guo+2018} sample (with $f_{LUV} > 5\%$), herein called the Guo+2018 sample. Guo+2018 was chosen because it includes a large (523 clumps) sample with $f_{LUV} \geq 5\%$, and because it spans a similar physical resolution to \textit{Clump Scout}: A typical SDSS g-band PSF-FWHM at $z \sim 0.05$ is $\sim$1.2 kpc, compared with $\sim$1.1-1.3 kpc for HST sources spanning $1<z<2.5$ for similar wavelengths.

For each clump in the Guo+2018 sample, we examined its flux fraction in CANDELS filter bands that were analogous to near UV and the SDSS $u$ band in the rest frame.\footnote{The filters selected for UV were F435W for $0.5<z<1$, F606W for $1<z<2$, and F775W for $2<z<3$, matching the scheme in \citet{Guo+2015}. We selected SDSS $u$ band analog filters to be closest to 3550\AA in the rest frame: F606W for $0.5<z<0.9$, F775W for $0.9<z<1.2$, F814W for $1.2<z<1.6$, F105W for $1.6<z<2.2$, and F125W for $2.2<z<3$.} We find that there is a strong correlation between $f_{Lu}$ and $f_{LUV}$ using these filters, with the median clump having $f_{Lu} = 0.86 f_{LUV}$. A total of 1,170 clumps from the Guo+2018 sample meet the $f_{LUV} > 8\%$ criterion, compared with 961 which meet $f_{Lu} > 8\%$, a reduction of $\sim18\%$.

\begin{figure}[t]
    \centering
    \includegraphics[width=\columnwidth]{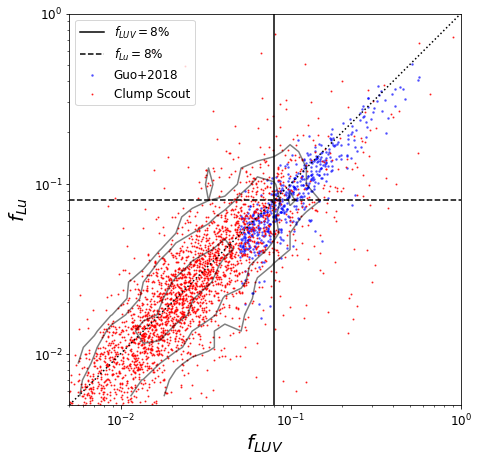}
    \caption{Comparison of $f_{LUV}$ (the clump-to-galaxy flux ratio in the near UV) and $f_{Lu}$ (the same in the SDSS $u$ band) in the Guo+2018 sample and the \textit{Clump Scout} sample. The y-axis traces $f_{Lu}$ , while the x-axis traces $f_{LUV}$ (the same in the near UV). Red points are used for \textit{Clump Scout} clumps while the gray contour lines are drawn to visualize their scatter. For Guo+2018 clumps, $f_{LUV}$ and $f_{Lu}$ have a strong linear correlation (Spearman rank correlation $r_s \approx 0.84$). We take the near UV fluxes for a sample of \textit{Clump Scout} galaxies from the GALEX survey. While we cannot measure the near UV fluxes of \textit{Clump Scout} clumps directly, we estimate it by assuming that local clumps have similar SEDs to clumps in the Guo+2018 sample and multiplying each clump's u-band flux by the UV-to-u ratio for a randomly selected Guo+2018 clump. The resulting distributions of $f_{LUV}$ and $f_{Lu}$ fall nearly along a 1-1 for both \textit{Clump Scout} and Guo+2018. $f_{Lu}$ is strongly correlated with $f_{LUV}$, though there is a large degree of scatter in the relation.}
    \label{fig:cs_vs_guo_relflux}
\end{figure}

We performed a similar experiment on clumps in \textit{Clump Scout}. Though we could not calculate $f_{LUV}$ directly since SDSS does not provide near UV data ($\sim 2500$\AA), we obtained near UV fluxes for local galaxies from the GALEX survey \citep{Martin+2005_GALEX}, using the cross-matched GALEX-SDSS catalog created by \cite{BianchiShiao2020}. We then assumed that local clumps matched the SED distribution of clumps from the Guo+2018 sample and multiplied each clump's $u$ band flux by the UV-to-$u$ ratio of a randomly selected Guo+2018 clump. The resulting values for $f_{Lu}$ and $f_{LUV}$ are plotted alongside the Guo+2018 values in Figure \ref{fig:cs_vs_guo_relflux}. For both groups, $f_{Lu}$ is a reasonably strong predictor of $f_{LUV}$. Performing the same experiment using the $g$ band rather than the $u$ band revealed that $f_{Lg}$ values are not as well correlated to $f_{LUV}$ and are typically much smaller (the median clump had $f_{Lg} = 0.73 f_{LUV}$).

Based on these experiments, we conclude that $f_{Lu}$ is the best available analog of $f_{LUV}$ for SDSS data, though it results in a slight underestimate of $f_{clumpy}$.

\subsection{Calculating $f_{clumpy}$}

\begin{table*}[]
    \centering
    \begin{tabular}{|l|l|l|l|}
        \hline
        & \multicolumn{1}{c|}{\textbf{Regular}} & \multicolumn{1}{c|}{\textbf{Extra (Examined)}} & \multicolumn{1}{c|}{\textbf{Total}} \\ \hhline{|=|=|=|=|}
        Parent sample & 53,613 & 171,472 (4,937) & 225,085 \\ \hline
        With sSFR $>$ 0.1 Gyr$^{-1}$ & 12,671 & 21,719 (1,015) & 34,390 \\ \hline
        With b/a ratio $>$ 0.3 & 12,142 & 20,572 (954) & 32,714 \\ \hhline{|=|=|=|=|}
        \textbf{All galaxies} & 2,640 & 3,290 (230) & 5,930 \\
        ($10^9 < M/M_\odot$, $z<0.035$) & & & \\ \hline
        \textbf{Low-mass galaxies} & 2,042 & 2,939 (118) & 4,981 \\
        ($10^9 < M/M_\odot < 10^{9.8}$, $z<0.035$) & & & \\ \hline
        \textbf{Medium-mass galaxies} & 1,741 & 1,554 (149) & 3,295 \\
        ($10^{9.8} < M/M_\odot < 10^{10.6}$, $z<0.05$) & & & \\ \hline
        \textbf{High-mass galaxies} & 601 & 868 (203) & 1,469 \\
        ($10^{10.6} < M/M_\odot < 10^{11.4}$, $z<0.09$) & & & \\ \hline
    \end{tabular}
    \caption{The cuts applied on the galaxy sample used for calculating $f_{clumpy}$. ``Regular'' galaxies were fully examined by \textit{Clump Scout} volunteers, while ``extra'' galaxies (with $f_{merger} \geq 0.5$ or $f_{featured} \leq 0.5$) were only partially examined (the number of examined galaxies is given in parentheses); the ``total'' column sums the regular and extra columns. Row 1 presents the parent sample of galaxies from Table \ref{tab:sample_selection}, while rows 2 and 3 enumerate the sample removing quiescent galaxies and edge-on galaxies respectively. The final four rows detail the four mass-binned samples used for calculating $f_{clumpy}$.}
    \label{tab:sample_selection_fclumpy}
\end{table*}

The calculation of $f_{clumpy}$ from the Clump Scout catalog has several steps. We calculate $f_{clumpy}$ in the full sample as well as in 3 distinct bins of galaxy mass, and the selection cuts and galaxy counts for each mass bin are detailed in Table \ref{tab:sample_selection_fclumpy}. Here, we detail the steps for calculating $f_{clumpy}$ in the broadest mass bin ($M > 10^9 M_\odot$), though the same process applies to every bin.

First, we isolate a star-forming, mass-complete sample of galaxies. SDSS is complete for galaxies down to $10^9 M_\odot$ at redshifts $z<0.035$. Therefore, beginning with the \textit{Clump Scout} parent sample defined in section \ref{sec:galaxy_selection}, we limit the sample to galaxies with specific star formation rate (sSFR) $> 10^{-1}$, $M > 10^9 M_\odot$, and $z < 0.035$. In addition, because clumps may be more difficult to detect in edge-on galaxies than face-on galaxies, we remove all galaxies for which the ratio of the galaxy's major to minor axis is less than 0.3, where this axis ratio is estimated by the SDSS exponential fit in the r-band (\texttt{expAB\_r} in the PhotoPrimary table). SDSS contains $N_{tot} =$ 5,930 galaxies passing these cuts.

We then combine the contributions from the ``regular'' \textit{Clump Scout} sample of 53,613 galaxies, and the ``extra'' sample of 4,937 galaxies used to extrapolate over all galaxies in SDSS that \textit{Clump Scout} did not directly examine. In total, $N_{reg} =$ 2,640 of 5,930 galaxies passing all cuts for $f_{clumpy}$ were examined in the regular sample.

Of the remaining 3,290 galaxies, a sample of 230 were examined by volunteers as part of the ``extra'' sample.  We use $N_{extra,samp}$ to refer to the size of the sample of these galaxies that volunteers examined directly, ie. $N_{extra,samp} = $ 230, and $N_{extra,tot}$ to refer to the size of the total population from which these galaxies were drawn, ie. $N_{extra,tot} = $ 3,290.

For each group, the observed clumpy fraction $f_{clumpy,obs}$ is calculated and the completeness correction from Section \ref{sec:fclumpy_completeness_corr} is applied. The corrected fraction $f_{clumpy,extra}^{corr}$ over the ``extra'' sample is then extrapolated over all SDSS galaxies not examined by \textit{Clump Scout}. This yields the total clumpy fraction:
\begin{equation}
f_{clumpy} = \frac{N_{reg}}{N_{tot}} f_{clumpy, reg} + \frac{N_{extra,tot}}{N_{tot}} f_{clumpy, extra}
\end{equation}
Here, $f_{clumpy,reg}$ and $f_{clumpy,extra}$ are the fraction of galaxies out of $N_{reg}$ and $N_{extra,samp}$ respectively that were estimated to be clumpy.

The sampling error is estimated separately on the ``regular'' and ``extra'' samples using the standard error formula on a proportion, taking the sample size $N$ to be the number of examined galaxies in the ``regular'' or ``extra'' group:
\begin{equation}
\begin{split}
\sigma_{clumpy,reg} &= \sqrt{\frac{f_{clumpy,reg} (1 - f_{clumpy,reg})}{N_{reg}}} \\
\sigma_{clumpy,extra} &= \sqrt{\frac{f_{clumpy,extra} (1 - f_{clumpy,extra})}{N_{extra,samp}}}
\end{split}
\end{equation}
We then scale these by their contribution to the total value of $f_{clumpy}$, and add them in quadrature to obtain
\begin{equation*}
\begin{split}
\sigma_{clumpy}^2 = &\Big( \frac{N_{reg}}{N_{tot}}\Big)^2 \sigma_{clumpy, reg}^2 \; + \\ &\Big( \frac{N_{extra,tot}}{N_{tot}} \Big)^2 \sigma_{clumpy, extra}^2
\end{split}
\end{equation*}

To estimate the total error on $f_{clumpy}$, we use a Monte Carlo method to include contributions from the uncertainty on clump fluxes and clump incompleteness as well as from sampling error. Over 100 trials, clump fluxes are allowed to randomly vary within a normal distribution defined by their estimated error values (see Section \ref{sec:flux_recovery}), while the clump completeness map is recalculated on each trial by the method described in Appendix \ref{app:completeness_calculation}. On each Monte Carlo trial, we include sampling error by calculating an initial value of $f_{clumpy}$, then reassigning it to a random value selected from $N(f_{clumpy}, \sigma_{clumpy})$. Of these sources of error, sampling error is by far the most significant: In a trial run where error contributions from clump flux error and the completeness correction were ignored, the error bars were typically within 10\% of their original values. The exception is $f_{clumpy,3\%}^{corr}$ whose error bars were $\sim$40\% lower when completeness and clump flux error contributions were ignored.

\subsection{Correcting $f_{clumpy}$ for incompleteness} \label{sec:fclumpy_completeness_corr}

Our estimate of $f_{clumpy}$ must also take into account the incompleteness of our clump catalog. We therefore use the following method for completeness-correcting the clumpy fraction of galaxies.

The end-goal of the $f_{clumpy}$ completeness correction is to calculate $P_{FN}$, the probability that a galaxy is a false negative for clumps. In other words, $P_{FN}$ is the probability that a given galaxy contains one or more clumps, but that none of its clumps were detected. To calculate this probability, we work in steps from the completeness estimates on individual clumps in the sample. We define $P_{rec,i}$ to be the recovery fraction of clump $i$. (Refer to Appendix \ref{app:completeness_calculation} for a full overview of how $P_{rec,i}$ is determined for each clump.)

We begin by calculating the recovery probability of a randomly selected clump within the sample, $P_{rec}$:
\begin{equation} \label{eq:prec}
    P_{rec} = \frac{n_{obs}}{n_{true}} \approx \frac{n_{obs}}{\sum_{i=1}^{n_{obs}} 1/P_{rec,i}}
\end{equation}

\begin{figure*}
    \centering
    \includegraphics[width=0.85\textwidth]{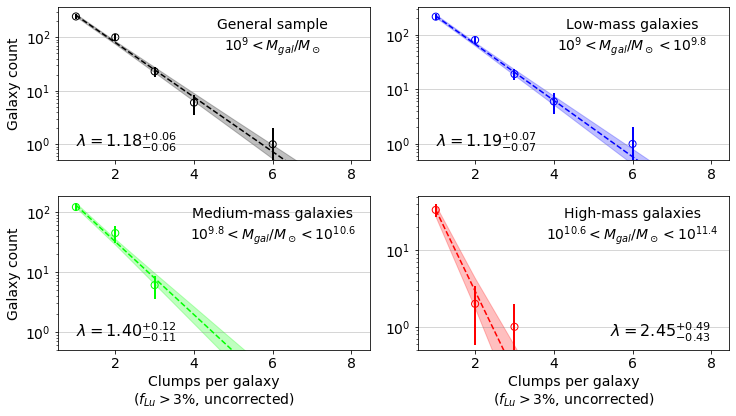}
    \caption{The distribution of clumps/galaxy in each of our 4 mass bins (the bins are described more fully in Table \ref{tab:fclumpy}). Dashed lines represent the best-fit exponential model to the data of the form $N_G = A e^{-\lambda n_c}$ for galaxy count $N_G$ and clumps per galaxy $n_c$. Error bars on this fit were estimated with Markov chain Monte Carlo sampling, and the shaded region around each line represents the 68\% confidence interval for the fit. The best-fit $\lambda$ value (with 68\% confidence interval) is also provided for each fit.}
    \label{fig:clump_count_distributions}
\end{figure*}

Next, we estimate the true distribution of clumps per clumpy galaxy, $P_{count}(n_c)$. To do so, we begin with a proposed distribution $P_{count}(n_c)$, then simulate 10,000 clumpy galaxies with this distribution. We then remove a fraction $P_{rec}$ of clumps at random to simulate the observed distribution, discarding galaxies with no observed clumps. The proposed distribution is adjusted until the mean number of clumps per galaxy in the simulated distribution closely matches the observed mean. For mathematical expediency, we used an exponential distribution to model $P_{count}(n_c)$. (It should be noted, though, that the observed distributions are also well-fit by Poisson distributions, so the exponential model does not necessarily have physical significance.) In Figure \ref{fig:clump_count_distributions}, we plot the observed distributions of $f_{Lu} > 3\%$ clumps per galaxy for each galaxy mass bin we examine along with the best-fit exponential model.

Given $P_{rec}$ and $P_{count}(n_c)$, $P_{FN}$ is given by
\begin{equation}
    P_{FN} = \sum_{n_c=1}^{\infty} P_{count}(n_c) \, (1-P_{rec})^{n_c}
\end{equation}
The $P_{FN}$ sum is dominated by the first few terms. For $f_{Lu} > 8\%$ clumps in our broadest galaxy mass bin ($M > 10^9 M_\odot$ with $z < 0.035$), we estimate $P_{rec} \approx 49.3\%$ and $\lambda \approx 0.85$; given these values, the first three terms of the $P_{FN}$ sum account for approx. 78\%, 92\% and 99\% of missed clumpy galaxies respectively.

Given the galaxy false-negative probability $P_{FN}$, the completeness of $f_{clumpy}$ is given by $(1-P_{FN})$. It is then straightforward to correct the clumpy fraction:
\begin{equation} \label{eq:fclumpy_corr}
    f_{clumpy}^{corr} = (1-P_{FN})^{-1} f_{clumpy}^{obs}
\end{equation}
$f_{clumpy}^{corr}$ should be taken as our estimate of the ``true'' value, as it accounts for galaxies whose clumps were not detected; however, we present both the observed and corrected values in our results.

\subsection{Results}

\begin{figure}[t]
    \centering
    \includegraphics[width=\columnwidth]{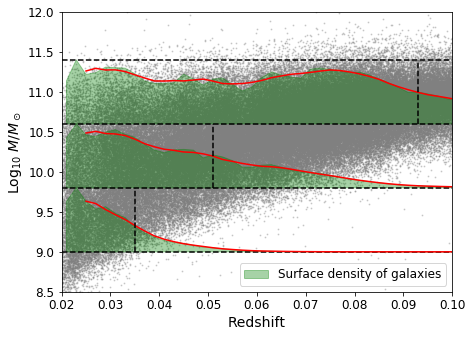}
    \caption{\textbf{Determining survey completeness limits for our mass bins.} Here we plot log mass vs. redshift for all galaxies examined by \textit{Galaxy Zoo 2} with spectroscopic redshifts (ie. the parent sample of \textit{Clump Scout}). Dashed horizontal lines demarcate the three primary mass bins used in our analysis. The green shaded regions represent the relative surface density of galaxies, while the red line traces the same with some added smoothing; assuming no significant cosmological evolution over this redshift range, the surface density should remain constant if there is no loss due to survey incompleteness. We draw a vertical dashed black line at the redshift where the surface density of galaxies first falls below 60\% of maximum, and we use this redshift to approximate the limit of a ``mass complete sample'' in each mass bin. Redshift limits are therefore drawn at $\sim 0.035$ for the lowest-mass bin, $\sim 0.05$ for the intermediate bin and $\sim 0.09$ for the highest mass bin.}
    \label{fig:mass_complete_sample}
\end{figure}

\begin{table*}[]
    \centering
    \begin{tabular}{|l||l|c|c|c||c|c|}
        \hline
        \multirow{2}{*}{} & & Galaxies & \multicolumn{2}{c||}{$f_{clumpy,8\%}$} & \multicolumn{2}{c|}{$f_{clumpy,3\%}$} \\
        sSFR $>$ 0.1 Gyr$^{-1}$ & $z_{max}$ &$(N_{tot})$ & Observed & Corrected & Observed & Corrected \\ \hline
        All masses & 0.035 & 5,930 & $1.65_{-0.22}^{+0.25}\%$ & $2.68_{-0.30}^{+0.33}\%$ & $6.54_{-0.40}^{+0.57}\%$ & $11.33_{-1.16}^{+0.89}\%$ \\ 
        & & & $\big(98_{-13}^{+14}\big)$ & $\big(159_{-18}^{+20}\big)$ & $\big(388_{-23}^{+34}\big)$ & $\big(672_{-69}^{+53}\big)$ \\ \hline
        
        Low mass & 0.035 & 4,981 & $1.67_{-0.19}^{+0.19}\%$ & $2.53_{-0.23}^{+0.21}\%$ & $6.57_{-0.61}^{+0.55}\%$ & $11.34_{-1.50}^{+1.09}\%$ \\ 
        ($10^9 < M / M_\odot < 10^{9.8}$) & & & $\big(83_{-9}^{+10}\big)$ & $\big(126_{-11}^{+11}\big)$ & $\big(328_{-31}^{+27}\big)$ & $\big(565_{-75}^{+54}\big)$ \\ \hline
        
        Medium mass & 0.05 & 3,295 & $1.53_{-0.49}^{+0.51}\%$ & $2.43_{-0.50}^{+0.57}\%$ & $5.55_{-0.74}^{+0.81}\%$ & $9.53_{-1.16}^{+1.14}\%$ \\ 
        ($10^{9.8} < M / M_\odot < 10^{10.6}$) & & & $\big(50_{-16}^{+17}\big)$ & $\big(80_{-16}^{+19}\big)$ & $\big(183_{-25}^{+27}\big)$ & $\big(314_{-38}^{+38}\big)$ \\ \hline
        
        High mass & 0.09 & 1,469 & $1.15_{-0.39}^{+0.35}\%$ & $1.95_{-0.40}^{+0.49}\%$ & $2.58_{-0.51}^{+0.44}\%$ & $4.88_{-0.68}^{+0.74}\%$ \\ 
        ($10^{10.6} < M / M_\odot < 10^{11.4}$) & & & $\big(17_{-6}^{+5}\big)$ & $\big(29_{-6}^{+7}\big)$ & $\big(38_{-8}^{+6}\big)$ & $\big(72_{-10}^{+11}\big)$ \\ \hline
    \end{tabular}
    \caption{Observed and corrected values for the clumpy fraction using the $f_{LUV} \geq 8\%$ and $f_{LUV} \geq 3\%$ criteria, divided by mass bin. Each cell reporting an estimate of $f_{clumpy}$ also reports the number of galaxies corresponding to this fraction, out of a possible total of $N_{tot}$. Note that the observed numbers refer to the sum of the clumpy galaxy count observed in the ``regular'' sample with the \textbf{extrapolated} count from the ``extra'' sample, so not all of these galaxies were directly observed.}
    \label{tab:fclumpy}
\end{table*}

\begin{figure}[h]
    \centering
    \includegraphics[width=\columnwidth]{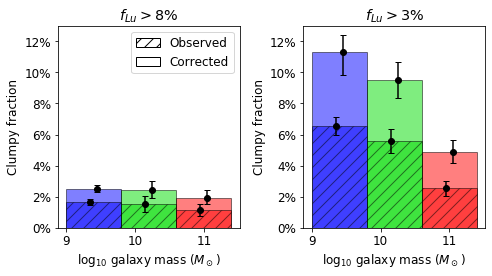}
    \caption{The value of $f_{clumpy}$ in three galaxy mass bins, under the criterion $f_{Lu} > 8\%$ (left) and $f_{Lu} > 3\%$ (right). The hatched bar represents the observed value, while the solid bar represents the completeness-corrected estimate.}
    \label{fig:fclumpy_mass_bins}
\end{figure}

Here, we present our results for $f_{clumpy,8\%}$ and $f_{clumpy,3\%}$ (the clumpy fraction using the thresholds $f_{Lu} > 8\%$ and $f_{Lu} > 3\%$ respectively), both overall and within several different galaxy mass bins. All of these numbers are collected in Table \ref{tab:fclumpy}.

Within the regular \textit{Clump Scout} sample ($N_{reg} =$ 2,640), we detected 85 galaxies with clumps passing $f_{Lu} > 8\%$; corrected for incompleteness, we estimate the true number to be $\sim$136. Within the extra sample ($N_{extra} =$ 230), we observed just 1 galaxy with clumps passing $f_{Lu} > 8\%$, and estimate the true number to be $\sim$2 correcting for incompleteness. In total we estimate that 157 of $N_{tot} = $5,930 galaxies have clumps (corrected for incompleteness), leading to an estimate $f_{clumpy,8\%}^{corr} = 2.68_{-0.30}^{+0.33}\%$.

We apply the same procedure to galaxies using the $f_{Lu} > 3\%$ cut, and observe 344 clumpy galaxies in the regular sample ($\sim$556 corrected for incompleteness) and 4 clumpy galaxies in the extra sample ($\sim$9 corrected for incompleteness). This yields a total estimate $f_{clumpy,3\%}^{corr} = 11.33_{-1.16}^{+0.89}\%$.

To characterize the local distribution of clumps, we also examine $f_{clumpy}$ in three different bins of galaxy mass: $9<log_{10}(M/M_\odot)<9.8$, $9.8<log_{10}(M/M_\odot)<10.6$, and $10.6<log_{10}(M/M_\odot)<11.4$; we refer to the galaxies in these bins as low-mass, medium-mass, and high-mass galaxies respectively. These match the mass bins over which \cite{Guo+2015} estimated the clumpy fraction. To obtain a complete galaxy sample, different redshift limits were applied to each bin: $z < 0.035$ for low-mass galaxies, $z < 0.05$ for medium-mass galaxies, and $z < 0.09$ for high-mass galaxies (see Figure \ref{fig:mass_complete_sample}).

Following the same procedure as for the overall clumpy fraction, we estimate that $f_{clumpy,8\%}^{corr}$ is $2.53_{-0.23}^{+0.21}\%$ for low-mass galaxies, $2.43_{-0.50}^{+0.57}\%$ for medium-mass galaxies, and $1.95_{-0.40}^{+0.49}\%$ for high-mass galaxies. A more complete list of statistics can be found in Table \ref{tab:fclumpy}; they are also plotted in Figure \ref{fig:fclumpy_mass_bins}. Example images of galaxies in each of the mass bins is shown in Figure \ref{fig:galaxy_images_massbins}.

\subsection{Comparisons to other studies}

\begin{figure*}[h!]
    \centering
    \includegraphics[width=\textwidth]{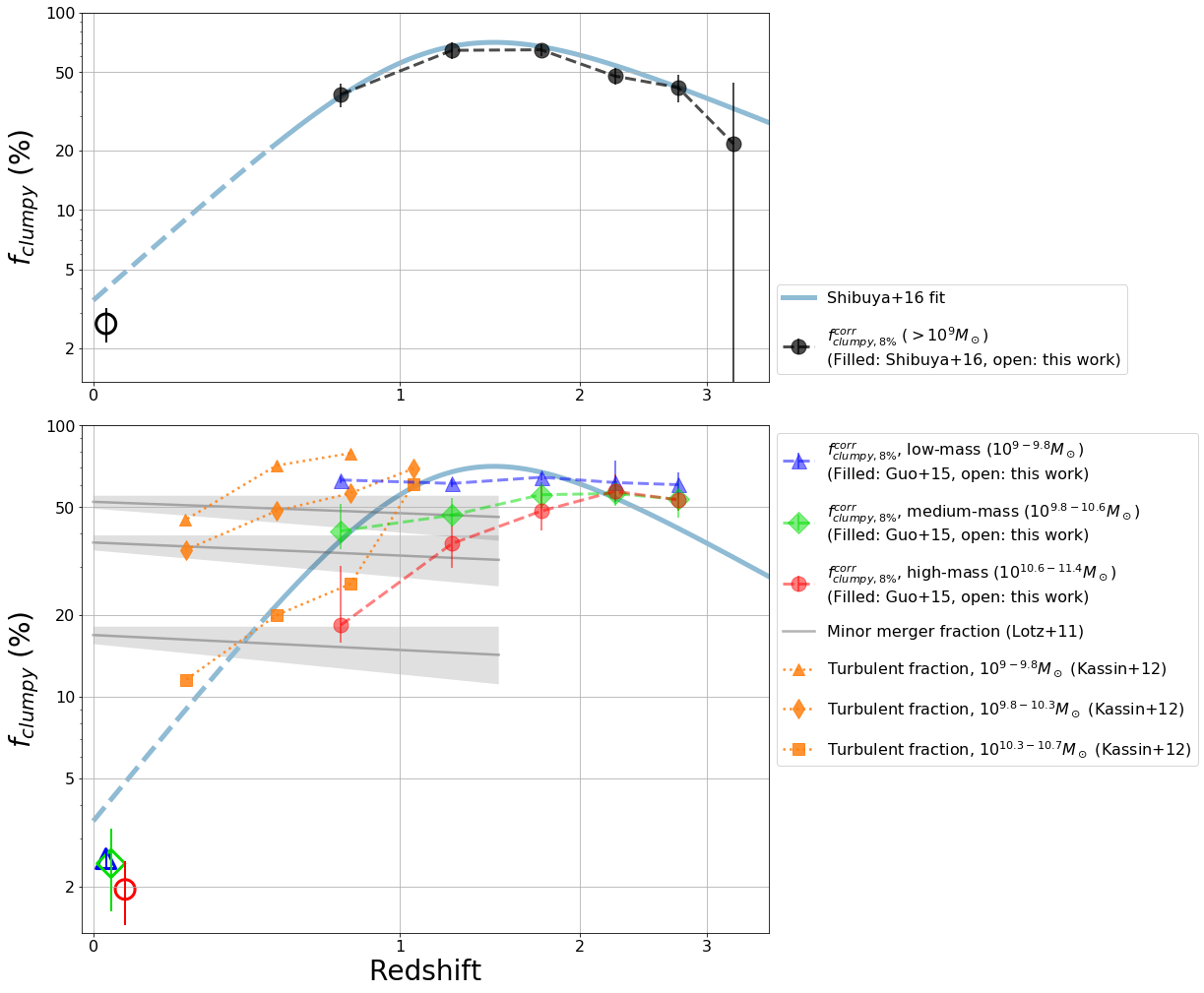}
    \caption{The value of $f_{clumpy,8\%}$ vs. redshift, as estimated by our study at $z\sim0$ and others at $z > 0.5$. \textbf{TOP:} $f_{clumpy}$ for galaxies of all masses ($> 10^9 M_\odot$). Our result is reasonably close to the model by \cite{Shibuya+2016} that was originally fit to their high-redshift $f_{clumpy}$ results. \textbf{BOTTOM:} $f_{clumpy}$ divided into mass bins. The mass bins and methods used in this work match closely with those used by \cite{Guo+2015} to estimate $f_{clumpy}$ at $z > 0.5$, but our clumpy fractions at $z\sim0$ are significantly lower than those at higher redshift. For comparison, we have plotted estimates of the minor merger fraction and observations of the ``turbulent fraction'' of galaxies between $0<z<1.5$. The minor merger fraction is modeled by \cite{Lotz+2011} and plotted here with observability timescales of 0.5, 1.25, and 2 Gyr (with gray error regions containing the best fit range). The turbulent fraction comes from kinematic observations by \cite{Kassin+2012} over three galaxy mass bins. We find that the turbulent fraction qualitatively matches the patterns observed in $f_{clumpy}$, ie. that it declines significantly over $1.5 < z < 0$ and that high-mass galaxies begin this decline the soonest. By comparison, we find that the minor merger fraction remains approximately constant over the same time period and is a poor tracer of $f_{clumpy}$ for any mass bin when including our results at $z \sim 0$.}
    \label{fig:fclumpy_comparison}
\end{figure*}

To place our estimates of $f_{clumpy,8\%}$ at $z < 0.1$ in context, we compare them with high-redshift ($z > 0.5$) results from other works, in particular \cite{Shibuya+2016} for galaxies of all masses $M > 10^9 M_\odot$ and \cite{Guo+2015} for galaxies in bins of low, medium, and high mass (matching the mass bins used in this paper). We plot these values in Figure \ref{fig:fclumpy_comparison}.

\cite{Shibuya+2016} found that the clumpy fraction peaks between redshifts 1-2 at a value of $> 50\%$ before declining over $z < 1$. To model this trend, they use a fit function with the same form as is commonly used to model the trend in the cosmic star formation rate density with redshift \citep{Madau+1996, Lilly+1996}. To compare with their results, we take their best fit model for $z$ vs. $f_{clumpy}$ and extend it beyond their data to $z\sim0$; this is plotted in Figure \ref{fig:fclumpy_comparison}. Their model predicts a value of $f_{clumpy,8\%} \sim 4\%$ at $z \sim 0$, which aligns closely with our result of $f_{clumpy,8\%} = 3.6 \pm 0.5\%$.

\cite{Guo+2015} observed a nearly constant $f_{clumpy,8\%}$ $\sim 50\%$ for low-mass galaxies over $0.5 < z < 3$, while our results indicate a value of $\sim 3\%$ at $z \sim 0$ for galaxies of the same mass. These results can be meaningfully compared, because the \textit{Clump Scout} results define $f_{clumpy,8\%}$ with a $f_{Lu} > 8\%$ cut that is similar to the $f_{LUV} > 8\%$ cut used by \cite{Guo+2015} (see Section \ref{sec:clump_def}). In addition, both probe a similar physical resolution: The physical resolution of CANDELS images is $\sim 1 kpc$ at all redshifts, compared to $\sim0.5$ to $\sim1.7$ kpc over the range $0.02<z<0.09$ for SDSS. We therefore expect that this drop in the clumpy fraction between $0 < z < 0.5$ is real and not merely the result of different identification methods.

There are few other studies that estimate $f_{clumpy}$ in the local universe. \cite{Murata+2014} studied clumpy galaxies selected from HST/ACS F814W imaging from the COSMOS field spanning redshifts $0.2 < z < 1$. The F814W filter approximately corresponds to the SDSS $r$ and $g$ bands over this redshift range and the nearest galaxies in this sample ($z\sim0.2$) are likely similar to \textit{Clump Scout} galaxies ($z<0.1$). The fraction of optically bright galaxies with multiple star-forming clumps was found to decrease from 0.35 at $z\sim1$ to 0.05 at $z\sim0.2$. While this decrease is qualitatively in line with our results, the definition of $f_{clumpy}$ used by \cite{Murata+2014} is significantly different than that used here: Rather than apply a relative flux criterion, clumpy galaxies were selected to have multiple star-forming clumps of comparable brightness. The low completeness of our sample prevents us from applying the \cite{Murata+2014} condition, as it requires the detection of at least 3 clumps per clumpy galaxy. \cite{Overzier+2009} also studied clumpy galaxies at $z<0.3$, but only studied a sample of 30 ``Lyman break analog'' galaxies with extremely high UV fluxes; the clumpy fraction obtained from this sample is not comparable to that of our broader sample.

\section{Discussion}

\subsection{Interpretation of $f_{clumpy}$}

\textbf{Given this paper’s focus on the fraction of clumpy galaxies, it is worth discussing exactly how this quantity is defined and how it should be used. $f_{clumpy}$ is a particularly good probe of trends in clumpiness across cosmic time: By controlling for galaxy mass and star-formation rate, we ensure that $f_{clumpy}$ is computed between groups of similar galaxies even at different redshifts. However, using a relative flux criterion ($f_{LUV} > 8\%$) to define clumps may select for very different sets of physical objects depending on galaxy mass. Naively, assuming a linear relation between mass and u-band luminosity in our galaxy sample, our lowest galaxy mass bin ($10^9 - 10^{9.8} M_\odot$) includes clumps that are $\sim2$ dex less massive than our highest mass bin ($10^{10.6} - 10^{11.4} M_\odot$). It is therefore not straightforward to compare $f_{clumpy}$ between bins of different galaxy mass. The validity of this comparison depends on the clump luminosity function: For example, if the clumpy luminosity function experiences an exponential cutoff \citep[e.g. as proposed by][]{Livermore+2012}, the relation between $f_{clumpy}$ and galaxy mass would depend the location of this exponential cutoff with galaxy mass. Therefore, while $f_{clumpy}$ can be compared for galaxies of similar mass across different redshifts, it is not straightforward to compare $f_{clumpy}$ between galaxies of different mass. The remainder of this discussion focuses on trends in $f_{clumpy}$ with redshift for this reason.}

\subsection{Physical implications of $f_{clumpy}$} \label{sec:fclumpy_implications}
A major motivation for determining $f_{clumpy}$ over large redshift ranges is to distinguish between different proposed modes of clump formation. There are two primary modes by which clumps are thought to form. In the in-situ mode, clumps form due to gas collapse within the host galaxy due to turbulent disk dynamics (ie. VDI). VDI is expected in galaxies that are actively accreting gas via ``cold-mode'' accretion, in which gas flows into the galaxy via smooth, cold streams. This accretion process adds kinetic energy to the disk and can drive the Toomre parameter below unity, making gas unstable to collapse \citep{DekelSariCeverino2009}. Alternately, in the ex-situ mode of formation, clumps originate as minor mergers: The clump forms as a satellite galaxy with its own dark matter component, only later merging with its host. It should be noted that clumps are short-lived structures on a cosmological scale: Simulations find that massive clumps in disk galaxies have a maximum lifetime of $\lesssim 500$ Myr, by which time they are slowed due to dynamical friction and have merged with their host's central bulge \citep{Bournaud+2014, Mandelker+2014}. Therefore, the presence of clumps indicates that the clump formation process is ongoing or recent, and $f_{clumpy}$ can effectively act as a tracer of galaxy behavior. It remains unclear which is the dominant formation process as  different processes may dominate different galaxy populations.

To determine the primary formation process of clumps (i.e. via in-situ or ex-situ formation), we can examine trends in the rate of VDI and the minor merger rate over cosmic time and compare these with trends in $f_{clumpy}$. In Figure \ref{fig:fclumpy_comparison}, we have plotted our estimate of $f_{clumpy,8\%}$ along with comparable estimates at higher redshift \citep{Guo+2015, Shibuya+2016} and estimates of the fraction of galaxies experiencing VDI and with observable signatures of minor mergers.

We use the minor merger rate estimate from \cite{Lotz+2011}, which was obtained by subtracting the number of galaxies with close pairs (major mergers) from the number with disturbed, uneven morphologies (major and minor mergers). The best-fit model to the minor merger fraction takes the form $f_{merg,minor} \propto T_{obs} (1+z)^\alpha$, with best-fit exponent $\alpha = -0.2 \pm 0.2$. The ``observability timescale'' refers to the time during which the host galaxy's morphology is measurably disturbed, which is dependent on the detection method (and distinct from the lifetime of an ex-situ clump formed during a merger). To represent the uncertainties in these parameters, we plot the best fit model for $T_{obs}$ values of 0.5, 1.25, and 2 Gyr over the fit range ($0 < z < 1.5$). In all cases, the minor merger rate rises or remains constant over the full redshift range due to the fit parameter $\alpha = -0.1 \pm 0.1$.

To determine the fraction of galaxies experiencing turbulence, we use measurements of galaxy kinematics from \cite{Kassin+2012}. These measurements reveal that galaxies of a wide range of masses ($10^{8-10.7} M_\odot$) tend to ``settle'' and become rotationally-dominated over the period $0< z <1.2$. Moreover, they find that the highest mass galaxies have the lowest fraction of turbulence at any epoch. The ``turbulent fraction'', defined as the fraction of galaxies for which $V_{circ} / \sigma_{gas} < 3$ (ie. the fraction of galaxies experiencing VDI), is plotted in Figure \ref{fig:fclumpy_comparison} for 3 of the mass bins examined by \cite{Kassin+2012}, spanning $10^{9-10.7} M_\odot$. All turbulent fractions decline over $0 < z < 1.2$, with higher-mass galaxies declining more quickly.

We then turn to trends in $f_{clumpy,8\%}$ and compare them to the trends in the two clump formation mechanisms (VDI and minor mergers) described above. Ignoring our low-redshift data for a moment, the data from \cite{Guo+2015} suggested that two different clump formation mechanisms may be dominant in high-mass galaxies and low-mass galaxies. The clumpy fraction for high-mass galaxies declines significantly with time over the span $0.5 < z < 3$ from $\sim55\%$ to $\sim15\%$, while for low-mass galaxies it remains constant at $f_{clumpy} \sim 60\%$ over the same time span. To explain this difference, it was suggested that the primary formation mechanism for clumps in high-mass galaxies may be VDI (in-situ) and trace the turbulent fraction over this time span, while those in low-mass galaxies form due to minor mergers (ex-situ) which are roughly stable over the same time span.

However, our low-redshift estimates of $f_{clumpy}$ challenge this two-mechanism formation model. For galaxies of all masses, we now observe a significant decline in $f_{clumpy,8\%}$ to $< 5\%$ at $z \sim 0$. Even assuming that our $z \sim 0$ estimates of $f_{clumpy,8\%}$ are too small by a factor of several, the observed fraction would still be far lower at $z \sim 0$ than at $z > 0.5$ in every mass bin. This result matches the conventional wisdom about clumps, i.e. that giant star-forming clumps are common at high redshift and rare locally. However, the roughly constant minor merger rate over the period $0 < z < 0.5$ is inconsistent with the significant low-redshift decline that we observe.

Instead, we suggest that in-situ, VDI-driven formation is the primary mode of clump formation in galaxies of all masses, at least over the redshift range $0 < z < 1.5$. The trends in galaxy turbulence over this time span match closely with the trends in $f_{clumpy}$: All galaxies show evidence of a decline in turbulence, with low-mass galaxies remaining turbulent the longest. The VDI-driven formation model provides a natural mechanism for the decline in $f_{clumpy}$, which is the decline in the cosmological rate of gas accretion by galaxies: As the availability of intergalactic gas decreases, so too do the rates of star-formation, turbulent dynamics, and clump formation \citep{DekelSariCeverino2009}.

Adding to this picture, smaller case studies have already provided limited evidence to link VDI to clumpiness directly. Studies of the kinematics of high-redshift clumpy galaxies find that they have turbulent morphologies \citep{Elmegreen+2009b, Genzel+2011}, with \cite{Genzel+2011} in particular noting that clumps appear in regions of the galaxy where the Toomre instability parameter is sub-unity. However, these studies examined spirals with stellar masses $\gtrsim 10^{10.6} M_\odot$, corresponding to the highest mass bin in our work; similar kinematic examination of galaxies with lower masses remains to be done. In total, the current body of evidence points to a picture of clump formation that is dominated by in-situ formation due to turbulent disk dynamics, though more concrete evidence would be needed to confirm this.

\section{Summary and conclusions}

In this work we present the largest-yet catalog of local star-forming clumps ($z \lesssim 0.1$), consisting of 14,341 clumps in 9,692 galaxies. The clumps were identified via the citizen science project \textit{Galaxy Zoo: Clump Scout} which asked volunteers to identify star-forming clumps in a sample of 58,614 galaxies selected from the parent sample \textit{Galaxy Zoo 2}. Consensus locations for these clumps are determined via an aggregation technique adapted from \cite{Branson+2017}. We estimate the completeness of our clump sample by comparing with a sample of simulated clumps identified via the same process.

The clump catalog generated by this work is versatile and can be used for many purposes. While this paper focused on estimating the clumpy fraction of galaxies, the catalog can also be used to answer other questions about clumps. In follow-up work, we intend to investigate the mass and age functions of these clumps using photometric SED fitting. Obtaining these statistics will permit comparison between the properties of clumps at low- and high-redshift, and they can be also used to directly test theories of clump formation and evolution which make predictions on the mass or age distribution of clumps. 

We define two different measures of the clumpy fraction of galaxies, $f_{clumpy,8\%}$ and $f_{clumpy,3\%}$, which measure the fraction of galaxies with at least one clump emitting at least 8\% and 3\% of galaxy flux in the $u$ band ($f_{Lu} \geq 8\%$ and $f_{Lu} \geq 3\%$) respectively. $f_{Lu}$ is found to be the closest analog to $f_{LUV}$ (the fraction of galaxy flux emitted in the near UV) available in SDSS data. $f_{clumpy,8\%}$ is presented because it has been used in the past for high-redshift studies of clumpy galaxies, while $f_{clumpy,3\%}$ is presented for comparison with future local studies. Both fractions are corrected for incompleteness. We find $f_{clumpy,8\%} \sim 3.6\%$ and $f_{clumpy,3\%} \sim 13.4\%$, with considerable variation over different mass bins.

Our low value of $f_{clumpy,8\%}$ is qualitatively in line with other low-redshift surveys \citep[e.g.][]{Murata+2014} though few are available. It is however much lower than the values of $f_{clumpy,8\%}$ estimated at high redshift \citep{Guo+2015, Shibuya+2016}. We suggest that the extreme decrease in clumpy morphology is not in line with minor-merger-driven clump formation (as suggested by \cite{Guo+2015} for low-mass galaxies) because the minor merger rate does not show similar change over this period \citep{Lotz+2011}. Instead, we suggest that a better tracer of $f_{clumpy,8\%}$ is the turbulent fraction of galaxies. \cite{Kassin+2012} observed a decline in turbulence for galaxies of all masses ($10^{8-10.7} M_\odot$), but in particular noted that larger galaxies settle quickly after $z \sim 1.2$ while less massive galaxies remain turbulent for a longer time, mimicking the trends in $f_{clumpy}$. In total, the current body of evidence supports a picture where clumps primarily form in-situ due to disk instability, though more observations are needed.

\section*{Acknowledgements}

This research is partially supported by the National Science Foundation under grant AST 1716602.

This material is based upon work supported by the National Aeronautics and Space Administration (NASA) under Grant No. HST-AR-15792.002-A.

This research made use of Montage. It is funded by the National Science Foundation under Grant Number ACI-1440620, and was previously funded by the National Aeronautics and Space Administration's Earth Science Technology Office, Computation Technologies Project, under Cooperative Agreement Number NCC5-626 between NASA and the California Institute of Technology.

This publication uses data generated via the Zooniverse.org platform, development of which is funded by generous support, including a Global Impact Award from Google, and by a grant from the Alfred P. Sloan Foundation.

\bibliography{clumpy}

\appendix

\section{Galaxy cutout creation}
\label{app:cutout_creation}

Here, we detail the process by which we created galaxy image cutouts. For each galaxy, we began by downloading the $i$, $r$, and $g$ band fields listed in the SDSS PhotoPrimary table for the galaxy. From each field, we selected a cutout centered on the target galaxy and scaled to 6 times the $r$ band 90\% Petrosian radius on each side. Finally, we projected images in our three image bands to a uniform coordinate system using the Montage library \citep{Jacob+2010} to avoid sub-pixel offsets in the RGB composite image. This coordinate system had a scale of 0.396''/pixel to match SDSS's native resolution. To create a color composite image, the $i$, $r$ and $g$ band cutouts were mapped to red, green and blue respectively, and color scaling was performed via the ``Lupton scaling'' technique described in \cite{Lupton+2004}. The particular scaling performed obeyed the equation
\begin{equation}
    x = \frac{\text{asinh}(Q (I/\text{band\_scaling} - \text{minimum})/\text{stretch})}{Q} \\
\end{equation}
where $I$ is the input intensity in a given image band (in counts) and $x$ the output intensity for display (ranging from 0 to 1). ``$Q$'', ``minimum'' and ``stretch'' were tunable parameters set to 7, 0, and 0.2 respectively, while ``band\_scaling'' is a per-band parameter set to 1.818, 1.176, and 0.7 for the $i$, $r$ and $g$ bands respectively. The result resembles SDSS's standard color balance with a significant emphasis on the $g$ band to emphasize star-forming regions. Finally, images were resized to a standard 400$\times$400 pixel scale.

\section{Calculating the completeness estimate on clumps}
\label{app:completeness_calculation}

In this appendix we describe the mathematical process used to calculate the completeness estimate of each clump identified by \textit{Clump Scout} volunteers.

We estimate the completeness of each recovered clump as a function of three measured properties: The clump's brightness (traced by $m_g^{clump}$), the clump's color (traced by $m_g^{clump} - m_r^{clump}$), and the clump's brightness relative to the diffuse background (traced by $m^{clump}_g - m^{background}_g$). The background magnitude $m^{background}_g$ measures the estimated background in a circular aperture with diameter 1 PSF-FWHM. The $g$ band was selected as the primary band because it traces the bluest wavelength of the filters used in our image cutouts ($i$, $r$ and $g$), and correlates most strongly with star-formation. Other properties, such as galactocentric distance, galaxy mass, redshift, or image resolution, were found to have minimal impact, possibly because they are correlated with the three properties already used.

\begin{figure}[t]
    \centering
    \includegraphics[width=\columnwidth]{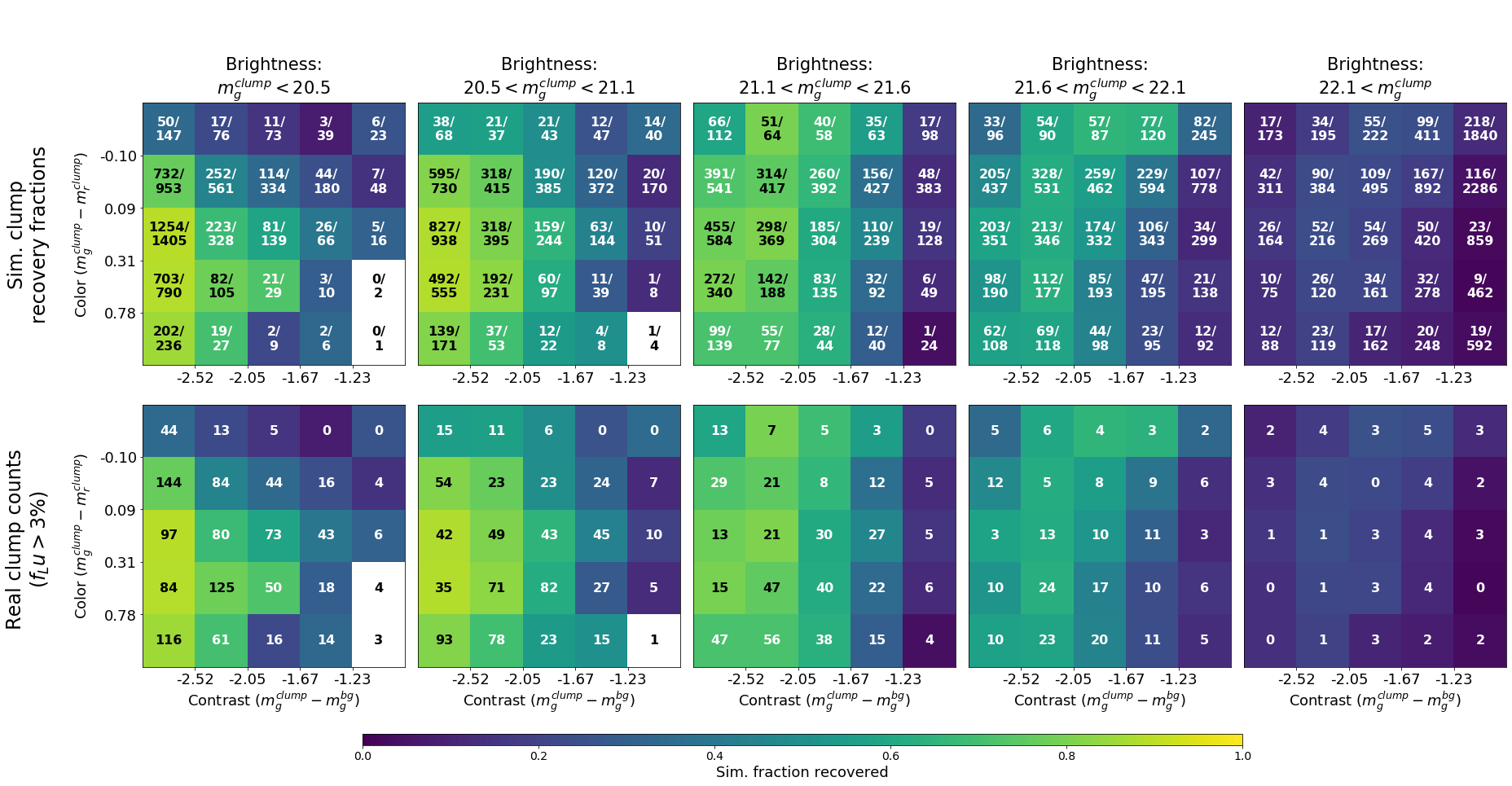}
    \caption{Demonstration of the simulated clump binning procedure. In each subplot, each shaded square represents a single bin in 3D space; the subplots divide clumps by their brightness, while the x- and y-axes trace the other two properties (contrast and color) that are included in the binning procedure. \textbf{TOP:} The recovery fraction of simulated clumps is printed in each bin, formatted as ``recovered/total''. \textbf{BOTTOM:} The number of real clumps with $f_{Lu} > 3\%$ is printed in each bin.}
    \label{fig:sim_clump_recovery_fractions}
\end{figure}

We then partitioned real clumps into bins based on their features $\bm x = \{x_1, x_2, x_3\}$, where
\begin{equation*}
\begin{split}
x_1 &= m_g^{clump} \\
x_2 &= m_g^{clump} - m_r^{clump} \\
x_3 &= m_g^{clump} - m_g^{background} \\
\end{split}
\end{equation*}
Partition edges were selected such that each feature $i$ was divided into partitions $k_i(x_i)$ containing equal number counts of real clumps. Partitions over all 3 features were then given by $\bm k (\bm x) = \{ k_1(x_1), k_2(x_2), k_3(x_3) \}$. (Note that since clump properties are not independent, the three-dimensional partitions $\bm k$ do \textbf{not} generally contain the same number counts of clumps.)

Next, we used simulated clumps to estimate the recovery fraction $f_{\text{recov}}(\bm k)$ in each partition. A naive estimate would be to equate $f_{\text{recov}}(\bm k)$ to the fraction of simulated clumps in bin $\bm k$ recovered by volunteers. However, this method is imprecise for bins with few recovered clumps and provides no simple method for estimating error. Instead we used a Bayesian method: We assumed that the likelihood distribution of $f_{\text{recov}}(\bm k)$ obeys a beta distribution defined by the number of recovered and missed clumps within $\bm k$. That is,
\begin{equation}
    P\big(f_{\text{recov}}(\bm k)\big) = B \big(n_{\text{recov}}(\bm k)+1, n_{\text{missed}}(\bm k)+1 \big)
\end{equation}
where $n_{\text{recov}}$ is the number of clumps in bin $\bm k$ that volunteers recovered and $n_{\text{missed}}$ is the number that they missed. We then took $f_{\text{recov}}(\bm k)$ to be the 50th percentile of the $P(f_{\text{recov}}(\bm k))$ distribution. (This method is described in \citet{Cameron2011}, and provides robust uncertainties for estimated fractions even when number count is low). Figure \ref{fig:sim_clump_recovery_fractions} displays the number of (total and recovered) simulated clumps in each bin, as well as the uncorrected distribution of recovered real clumps over these bins.

Finally, we interpolated the discrete map $f_{\text{recov}}(\bm k)$ to a continuous map over all clump properties, $f_{\text{recov}}(\bm x)$. We began with a grid of values for $f_{\text{recov}}(\bm x)$ by assuming that $f_{\text{recov}}(\bm k) = f_{\text{recov}}(\bm x_{50}(\bm k))$, where $\bm x_{50}(\bm k))$ is the median value of $\bm x$ for simulated clumps in partition $\bm k$. We then linearly interpolated over this grid to obtain the continuous map. Clumps falling outside of the interpolation grid were assigned $f_{\text{recov}}(\bm k(\bm x))$, ie. the value for their bin. 0.4\% of real clumps fell in bins with fewer than 5 simulated clumps; these clumps were discarded for further analysis.

Once the map $f_{recov}(\bm x)$ is established, we can estimate the specific completeness estimate for each clump in our sample: For clump $i$ with estimated properties $\bm x(i)$, we define its estimated completeness $P_{rec,i}$ as $P_{rec,i} = f_{recov} (\bm x(i))$. These $P_{rec,i}$ values can then be used to correct $f_{clumpy}$ for incompleteness, as explained in Section \ref{sec:fclumpy_completeness_corr}.

This method of estimating completeness also allows us to easily estimate the uncertainty on the estimate using a Monte Carlo method. Over 100 trials, we allowed the fraction of recovered clumps in each bin ($f_{\text{recov}}(\bm k)$) to vary randomly over its estimated distribution $P(f_{\text{recov}}(\bm k))$, rather than taking the 50th percentile value. Repeated trials yielded an approximate distribution on $f_{recov(x)}$.

\begin{figure}
    \centering
    \includegraphics[width=0.85\columnwidth]{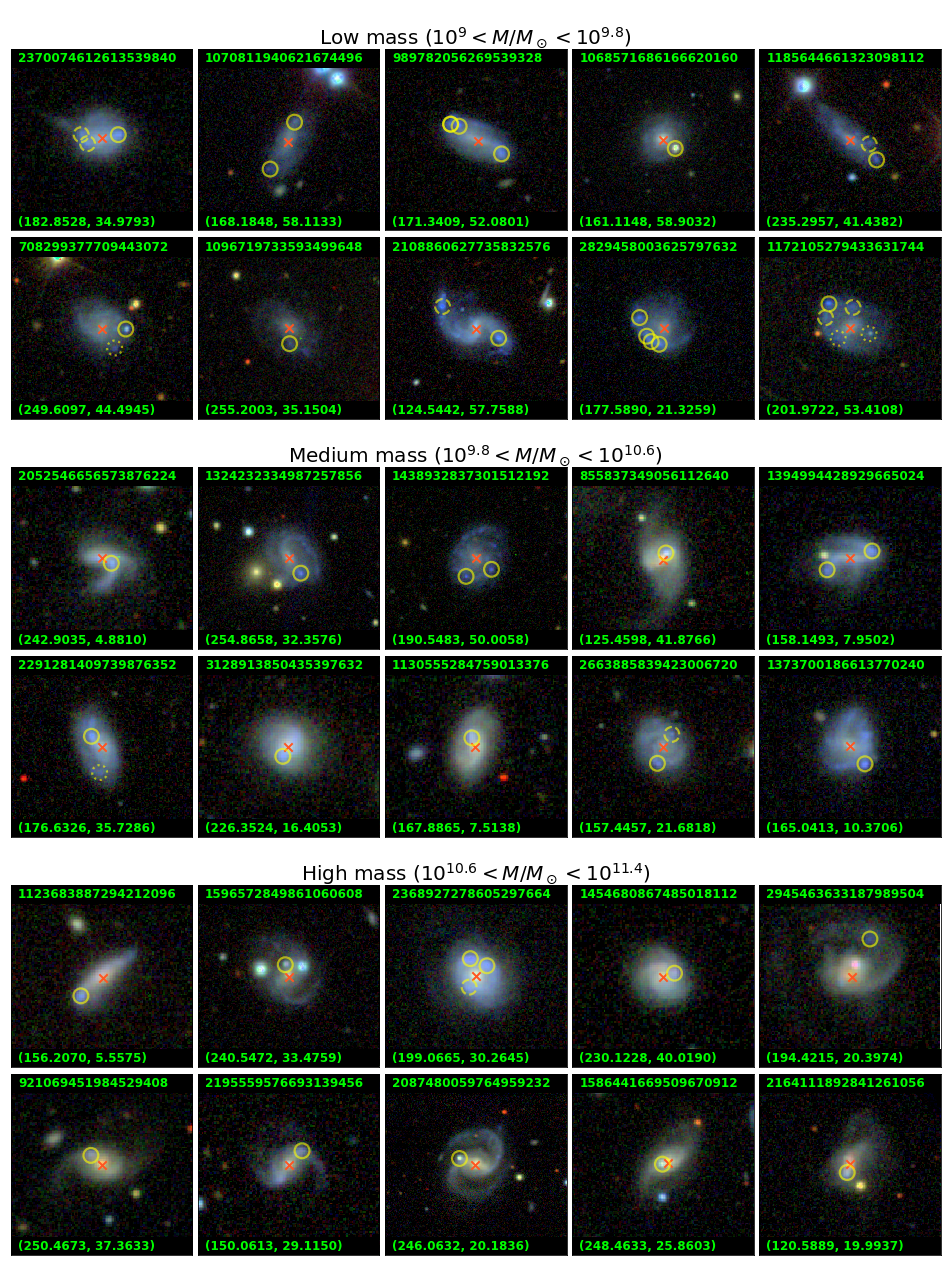}
    \caption{A sample of galaxies, each containing at least one clump with $f_{Lu} > 8\%$. Ten galaxies are presented from each of the low-, medium-, and high-mass bins; the SDSS specobjid of each galaxy is provided at the top of each image, while the RA and dec values are provided in parentheses at the bottom. Each identified clump is circled in yellow, while the central bulge locations identified by volunteers are identified by red crosses. Dotted circles indicate clumps with $f_{Lu} < 3\%$, dashed circles indicate those with $3\% < f_{Lu} < 8\%$, and solid circles those with $f_{Lu} > 8\%$}
    \label{fig:galaxy_images_massbins}
\end{figure}

\begin{table*}[]
    \centering
    \begin{tabular}{|l l l l|}
    \hline
    Column & Name & Note & Reference \\
    \hline \hline
    1 & ID & SDSS DR8 object ID & SDSS \\
    2 & Clump index & One-indexed & -- \\
    3 & Galaxy sample & Regular=0, Extra=1 & -- \\
    \hline 
    Galaxy properties & & & \\
    \hline
    4 & Redshift & & SDSS \\
    5-6 & Galaxy R.A., Dec. & J2000, deg & SDSS \\
    7 & Image r-band PSF-FWHM & arcsec & SDSS \\
    8-9 & Galaxy u-band flux and error & $\mu$Jy & SDSS \\
    10-11 & Galaxy g-band flux and error & $\mu$Jy & SDSS \\
    12-13 & Galaxy r-band flux and error & $\mu$Jy & SDSS \\
    14-15 & Galaxy i-band flux and error & $\mu$Jy & SDSS \\
    16-17 & Galaxy z-band flux and error & $\mu$Jy & SDSS \\
    18-19 & Galaxy near-UV flux and error & $\mu$Jy& GALEX \\
    20 & Galaxy r-band r$_{\text{eff}}$ & arcsec & SDSS \\
    21 & Galaxy log(M$^*$) & log(M$_\odot$) & MPA-JHU \\
    22 & Galaxy log(SFR) & log(M$_\odot$ Gyr$^{-1}$) & MPA-JHU \\
    23 & Galaxy log(sSFR) & log(Gyr$^{-1}$) & MPA-JHU \\
    \hline
    Clump properties & & & \\
    \hline
    24-25 & Clump R.A., Dec. & J2000, deg & \S\ref{sec:agg_method} \\
    26 & Clump offset & Normalized by galaxy r-band r$_{\text{eff}}$ & \S\ref{sec:agg_method} \\
    27 & Clump u-band flux and error & $\mu$Jy & \S\ref{sec:flux_recovery} \\ 
    28 & Clump g-band flux and error & $\mu$Jy & \S\ref{sec:flux_recovery} \\ 
    29 & Clump r-band flux and error & $\mu$Jy & \S\ref{sec:flux_recovery} \\ 
    30 & Clump i-band flux and error & $\mu$Jy & \S\ref{sec:flux_recovery} \\
    31 & Clump z-band flux and error & $\mu$Jy & \S\ref{sec:flux_recovery} \\
    32 & Background u-band flux and error & $\mu$Jy / arcsec$^2$ & \S\ref{sec:flux_recovery} \\
    33 & Background g-band flux and error & $\mu$Jy / arcsec$^2$ & \S\ref{sec:flux_recovery} \\
    34 & Background r-band flux and error & $\mu$Jy / arcsec$^2$ & \S\ref{sec:flux_recovery} \\
    35 & Background i-band flux and error & $\mu$Jy / arcsec$^2$ & \S\ref{sec:flux_recovery} \\
    36 & Background z-band flux and error & $\mu$Jy / arcsec$^2$ & \S\ref{sec:flux_recovery} \\
    37 & Est. clump near-UV flux and error & $\mu$Jy & \S\ref{sec:clump_def} \\
    38 & Est. clump/galaxy near-UV flux ratio & & \S\ref{sec:clump_def} \\
    39 & Volunteer annotation fraction & & \S\ref{sec:agg_method} \\
    40 & Volunteer unusual fraction & & \S\ref{sec:agg_method} \\
    41 & Unusual flag & & \S\ref{sec:catalog_use} \\
    42 & Completeness estimate & & \S\ref{sec:flux_recovery} \\
    
    \hline
    \end{tabular}
    \caption{\textbf{Clump Scout catalog description.} References abbreviated as ``SDSS'' refer to the SDSS DR15 catalog \citep{York+2000_SDSS_legacy}; those abbreviated as ``GALEX'' refer to the GALEX All-sky Imaging Survey \citep{Martin+2005_GALEX}, with SDSS cross-matching performed by \cite{BianchiShiao2020}; those abbreviated as ``MPA-JHU'' refer to the value-added catalog released with SDSS, based on work by \cite{Kauffmann+2003_MPA-JHU} and \cite{Brinchmann+2004_MPA-JHU}.}
    \label{tab:catalog}
\end{table*}

\end{document}